\documentclass[]{aa520}

\usepackage{graphicx}
\usepackage{natbib}
\usepackage{txfonts}

\newcommand{\cmd}       {\mathrm{\,cm^{-3}}}
\newcommand{\erg}       {\mathrm{\,erg}}
\newcommand{\gyr}       {\mathrm{\,Gyr}}
\newcommand{\kelvin}    {\mathrm{\,K}}
\newcommand{\kms}       {\mathrm{\,km\,s^{-1}}}
\newcommand{\kpc}       {\mathrm{\,kpc}}
\newcommand{\msun}      {\mathrm{\,M}_{\sun}}
\newcommand{\msunpcd}   {\mathrm{\,M_\odot \,pc^{-3}}}
\newcommand{\msunpcdmyr}{\mathrm{\,M_\odot \,pc^{-3}\,Myr^{-1}}}
\newcommand{\msunpcz}   {\mathrm{\,M_\odot \,pc^{-2}}}
\newcommand{\msunyr}    {\mathrm{\,M_\odot\,yr^{-1}}}
\newcommand{\myr}       {\mathrm{\,Myr} }
\newcommand{\pcmyr}     {\mathrm{\,pc\,Myr^{-1}}}
\newcommand{\pczmyrz}   {\mathrm{\,pc^2\,Myr^{-2}}}
\newcommand{\press}     { {P_\mathrm{e}/\mathrm{k} \over 10^4
                          \mathrm{K}/\mathrm{cm}^3} }
\newcommand{\pc}        {\mathrm{\,pc}}
\newcommand{\vrot}      {\mathrm{\,v_{rot}}}
\newcommand{\yr}        {\mathrm{\,yr}}

\newcommand{\ccoll}     {c_\mathrm{coll}}
\newcommand{\csf}       {c_\mathrm{sf}}
\newcommand{\ekcld}     {E_{\mathrm{cld}}}
\newcommand{\ekicm}     {E_{\mathrm{icm}}}
\newcommand{\esn}       {E_{\mathrm{sn}}}
\newcommand{\etacld}    {\eta_{\mathrm{cld}}}
\newcommand{\etacoll}   {\eta_{\mathrm{coll}}}
\newcommand{\etacond}   {\eta_{\mathrm{cond}}}
\newcommand{\etacool}   {\eta_{\mathrm{cool}}}
\newcommand{\etaevap}   {\eta_{\mathrm{evap}}}
\newcommand{\etaicm}    {\eta_{\mathrm{icm}}}
\newcommand{\etasf}     {\eta_\mathrm{sf}}
\newcommand{\LCDM}      {$\Lambda$CDM }
\newcommand{\lcool}     {\Lambda(T_{\mathrm{icm}},Z_{\mathrm{icm}})}

\newcommand{\lymanalpha}{Ly$\alpha$ }
\newcommand{\rhcld}     {\rho_\mathrm{cld}}
\newcommand{\rhcon}     {\dot \rho_\mathrm{cond}}
\newcommand{\rhicm}     {\rho_\mathrm{icm}}
\newcommand{\rheva}     {\dot \rho_\mathrm{evap}}
\newcommand{\SFR}       {\dot \rho_\mathrm{sf}}
\newcommand{\sicld}     {\sigma_\mathrm{cld}}
\newcommand{\tdyn}      {t_\mathrm{dyn}}
\newcommand{\teicm}     {T_\mathrm{icm}}
\newcommand{\tcool}     {t_\mathrm{cool}}
\newcommand{\tsf}       {t_\mathrm{sf}}

\def\spose#1{\hbox to 0pt{#1\hss}}
\def\lta{\mathrel{\spose{\lower 3pt\hbox{$\mathchar"218$}}
     \raise 2.0pt\hbox{$\mathchar"13C$}}}

\begin{document}

  \title{The Formation of a Disk Galaxy \\ within a Growing Dark Halo}

  \author{Markus Samland \inst{1} \and Ortwin E.\ Gerhard \inst{1}}

  \titlerunning{Formation of a Disk Galaxy}

  \institute{Astronomisches Institut der Universit\"at Basel,
             Venusstrasse 7, CH-4102 Binningen}

  \offprints{ \\ M.Samland,\email{samland@astro.unibas.ch}}

  \date{Received \dots / Accepted \dots}
 
  \abstract{We present a dynamical model for the formation and
    evolution of a massive disk galaxy, within a growing dark halo
    whose mass evolves according to cosmological simulations of
    structure formation. The galactic evolution is simulated with a
    new three-dimensional chemo-dynamical code, including dark matter,
    stars and a multi-phase ISM. The simulations start at redshift $z
    = 4.85$ with a small dark halo in a \LCDM universe and we follow
    the evolution until the present epoch. The energy release by
    massive stars and supernovae prevents a rapid collapse of the
    baryonic matter and delays the maximum star formation until
    redshift $z \approx 1$. The metal enrichment history in this model
    is broadly consistent with the evolution of [Zn/H] in damped
    \lymanalpha systems. The galaxy forms radially from inside-out and
    vertically from halo to disk. As a function of metallicity, we
    have described a sequence of populations, reminiscent of the
    extreme halo, inner halo, metal-poor thick disk, thick disk, thin
    disk and inner bulge in the Milky Way. The first galactic
    component that forms is the halo, followed by the bulge, the
    disk-halo transition region, and the disk. At redshift $z \approx
    1$, a bar begins to form which later turns into a triaxial
    bulge. Despite the still idealized model, the final galaxy
    resembles present-day disk galaxies in many aspects. The bulge in
    the model consists of at least two stellar subpopulations, an
    early collapse population and a population that formed later in
    the bar. The initial metallicity gradients in the disk are later
    smoothed out by large scale gas motions induced by the bar. There
    is a pronounced deficiency of low-metallicity disk stars due to
    pre-enrichment of the disk ISM with metal-rich gas from the bulge
    and inner disk (''G-dwarf problem''). The mean rotation and the
    distribution of orbital eccentricities for all stars as a function
    of metallicity are not very different from those observed in the
    solar neighbourhood, showing that early homogeneous collapse
    models are oversimplified.  The approach presented here provides a
    detailed description of the formation and evolution of an isolated
    disk galaxy in a \LCDM universe, yielding new information about
    the kinematical and chemical history of the stars and the
    interstellar medium, but also about the evolution of the
    luminosity, the colours and the morphology of disk galaxies with
    redshift. }

  \maketitle

  \section{Introduction} 

    During the last decade, significant progress has been made in
    understanding cosmic structure formation and galactic evolution.
    With high-resolution cosmological simulations, the formation of
    dark halos in different cosmologies has been studied in detail
    \citep[e.g.\,] []{navarro_96, moore_98, colin_00, yoshida_00,
    avila_01, jenkins_01, klypin_01} to determine the large-scale mass
    distribution, the halo merging histories, the structural
    parameters of the dark halos, and finally the formation of
    galaxies inside these dark halos \citep{steinmetz_95a, kay_00,
    navarro_00b, mosconi_01, pearce_01}.
 
    In many respects, these simulations are in agreement with
    observations, but there remain a number of difficulties. Firstly,
    the universal cuspy halo density profiles \citep{navarro_97,
    moore_99, fukushige_01}, while in agreement with observations of
    galaxy clusters, are in apparent disagreement with the flat dark
    matter density distributions inferred from observations in the
    centres of galaxies \citep{salucci_00, blais_01, borriello_01,
    deblok_01}. Secondly, the specific angular momenta and
    scale lengths of the disk galaxies in the cosmological simulations
    are too small compared to real galaxies \citep{navarro_00b}. This
    may be partially a numerical problem of SPH simulations, but is
    more likely due to overly efficient cooling and strong angular
    momentum transport from the baryons to the dark halos during
    merger events \citep{navarro_91}. \citet{thacker_01} and also
    \citet{sommerlarsen_99} showed the angular momentum problem in CDM
    simulations may be resolved by the inclusion of feedback
    processes.  Thirdly, the cold dark matter (CDM) simulations
    predict a large number of satellite galaxies in the neighbourhood
    of giant galaxies like the Milky Way Galaxy, but this is not
    observed \citep{moore_99}.
 
    The cosmological simulations are used to describe the large-scale
    evolution of the dark matter distribution from the primordial
    density fluctuations to the time when the dark halos form. It is
    inevitable in such simulations that structures on galactic and
    sub-galactic scales are not well-resolved. Furthermore, it is
    difficult to incorporate the detailed physics of the baryonic
    component, that is of the multi-phase interstellar medium (ISM),
    and the feedback processes between stars and ISM. These are
    serious drawbacks, because feedback from stars can prevent a
    proto-galactic cloud from rapid collapse and it can trigger large
    scale gas motions \citep{samland_97}. Both alter the galactic
    formation process significantly.
 
    Complementary ways to circumvent at least some of these problems
    are (i) to combine the large-scale simulations with
    semi-analytical galaxy models \citep{guiderdoni_98, somerville_99,
    boissier_00, cole_00, kauffmann_00}, or (ii) to simulate the
    formation of only single galaxies or small galactic groups using
    either cosmological initial conditions or a simplified, but
    cosmologically motivated model of the dark matter background
    \citep{navarro_97a, berczik_99, hultman_99, sommerlarsen_99,
    bekki_01, thacker_01, williams_01}. While the semi-analytical
    models are advantageous for investigating the global properties of
    galaxy samples, the small-scale dynamical models provide
    information about the detailed structure of galaxies, the
    kinematics of the stellar populations and the ISM, and the star
    formation histories. The dynamical models are called
    chemo-dynamical \citep{theis_92, samland_97, bekki_98,berczik_99,
    williams_01}, if they include different stellar populations, a
    multi-phase ISM, and an interaction network that describes the
    mass, momentum and energy transfer between these components.
 
    The aim of the present work is to investigate how a large disk
    galaxy forms inside a growing dark halo in a realistic \LCDM
    cosmogony. We present the results of a new three-dimensional
    model, including dark matter, stars, and a multi-phase ISM. The
    mass infall at the boundaries of the simulated volume is taken
    from cosmological simulations \citep[see
    VIRGO-GIF-project;][]{kauffmann_99}. We use a total dark matter
    mass of $1.5 \times 10^{12} \msun$, a total baryonic mass of $3
    \times 10^{11} \msun$, a spin parameter $\lambda = 0.05$, and an
    angular momentum profile similar to the universal profile found by
    \citet{bullock_01}. Different from \citet{bekki_98, berczik_99}
    and \citet{williams_01}, we use an Eulerian grid-code for the ISM,
    which we believe is better suited to describe the multi-phase
    structure of the ISM and the feedback processes, especially in
    low-density regions. The final galaxy model provides densities,
    velocities, velocity dispersions, chemical abundances, and ages
    for 614\,500 stellar particles, and temperatures, chemical
    abundances, densities, velocities, and pressures of the ISM. The
    goal of the project is a self-consistent model for the formation
    and evolution of large disk galaxies, which can be used to
    understand observations of young galaxies as well as data on
    stellar populations in the Milky Way.
 
    The paper is organized as follows. Section \ref{observations}
    summarizes some observational material relating to the formation
    and evolution of galaxies. Section \ref{model} describes our
    multi-phase ISM model and the underlying dark matter halo
    model. In Section \ref{history} we outline the formation history
    of the model disk galaxy. The properties of the galactic halo,
    bulge and disk components are discussed in Section
    \ref{components}. A summary and a short outlook conclude the paper
    (Section \ref{conclusions}).

  \section{Observations of the galaxy formation process}
  \label{observations}
 
    Galaxy evolution models can give a simplified description of the
    evolution of real galaxies and can make predictions that can be
    tested by observations. The high redshift observations show that
    the first proto-galactic structures form at redshift $z > 3$
    \citep[e.g.\,][]{clements_96, dey_98, steidel_99, manning_00}.
    Surveys of high redshift galaxies reveal a wide range of galactic
    morphologies with considerable substructure and clumpiness
    \citep{pentericci_01}. At redshift $z \approx 1-2$, the formation
    of galactic components (spheroids, halos, bulges and disks) seems
    to take place \citep{lilly_98, kajisawa_00}. However, the spatial
    resolution of these high and intermediate redshift surveys
    \citep[e.g.\,][]{abraham_99, vandenbergh_00, ellis_01,
    menanteau_01} is still too low to measure more than global
    galactic properties (e.g.\, asymmetry and concentration
    parameters). There are indications that most of the massive
    galaxies form before or around $z = 1$ \citep{brinchmann_00}, and
    especially that the giant ellipticals form at redshifts $z > 1$
    \citep{best_98}, \citep[but see][]{vandokkum_99}, but not before
    $z = 2.4$ \citep{pentericci_01}. There is a general trend of early
    type galaxies being older and having higher peak star formation
    rates than late type galaxies \citep{sandage_86}. These general
    star formation histories seem to be overlapped with star formation
    bursts, triggered by infalling material or interactions with other
    galaxies.
 
    Since we can observe galactic sub-structures down to the stellar
    scale only in our Galaxy, the Milky Way observations are a
    cornerstone for the understanding of the formation and evolution
    of all spiral galaxies. From studies of Galactic stars we know
    that the halo is old and that the disk and bulge contain a mixture
    of old and young stars. This raises the question in what sequence
    the galactic components halo, bulge and disk formed, and what was
    the star formation history. Determinations of the star formation
    history of the Milky Way and other nearby galaxies were attempted
    by \citet{bell_00, hernandez_00} and \citet{rocha-pinto_00}, based
    on star counts, broad band colours, or chemical compositions of
    stars.
 
    The chemical and kinematical data from galactic halo stars have
    been one of the starting points for different scenarios for the
    formation and evolution of galaxies. \citet{eggen_62} were the
    first to propose a monolithic collapse galactic formation model in
    which the Milky Way forms rapidly out of a collapsing
    proto-galactic cloud. This was based on data for galactic halo
    stars which are now known to have been plagued by selection
    effects \citep{chiba_00}. Later, based on observations of halo
    globular clusters, \citet{searle_78} proposed a new model in which
    the Milky Way forms out of merging lumps and fragments which all
    have their individual star formation histories. The contrasting
    nature of these two models and the proposed observable differences
    (e.g.\, existence or non-existence of metallicity gradients, the
    expected age spread for halo objects) seemed to offer a simple way
    to deduce the formation history of disk galaxies from Milky Way
    observations. However, new data especially from proper motion
    catalogues, lead to the conclusion that at least the halo of the
    Milky Way formed neither by a monolithic collapse nor by a pure
    merging process \citep{sandage_90, unavane_96, chiba_00}. In the
    modern hybrid scenarios, a part of the Galactic halo formed during
    a dissipative process (collapse) and another part shows signatures
    of merging events. For example, \citet{chiba_00} proposed a
    scenario in which \emph{``the outer halo is made up from merging
    and/or accretion of sub-galactic objects, such as dwarf-type
    satellite galaxies, whereas the inner part of the halo has
    undergone a dissipative contraction on relatively short time
    scales''}. This is consistent with the findings of
    \citet{helmi_99}, who found that 10\% of the metal-poor stars in
    the outer halo of the Milky Way are aligned in a single coherent
    structure, which they identify as the remnant of a disrupted dwarf
    galaxy. These hybrid scenarios are, in the general outline, also
    consistent with the formation of galaxies in a hierarchical
    universe \citep{chiba_00}.

  \section{The model}
  \label{model}
 
    The formation and evolution of a galaxy is influenced by the
    cosmological initial conditions and the environment, but also by
    internal feedback processes. Heating by supernovae (SNe),
    dissipation, radiative cooling, as well as in- and outflows
    influence the evolution both locally and globally. Our new
    three-dimensional galactic model takes into account the dynamics
    of stars and the different phases of the ISM, as well as the
    processes (``chemistry'') which connect the ISM and the stars.
    These baryonic components are embedded in a collision-free dark
    matter halo taken from simulations of cosmological structure
    formation. The description of the ISM is based on that used in the
    two-dimensional chemo-dynamical models of \citet{samland_96,
    samland_97}. The following subsections contain a brief outline of
    our model.

  \subsection{The dark halo}
  \label{darkhalo}
 
    Cosmological simulations \citep[e.g.\,][]{navarro_95, moore_99,
    pearce_99} show that in a hierarchical universe galaxies form out
    of a few large fragments with an additional slow accretion of
    mass. Thus each galaxy has its own formation history and
    characteristic shape, depending on the masses, angular momenta and
    merging histories of the fragments. Yet the formation of the CDM
    halos can be described by some simple ''universal'' formulae.
    \citet{wechsler_02} and \citet{vandenbosch_02} calculated the
    mass accretion history, \citet{bullock_01} the angular momentum
    distribution, \citet{navarro_97} and others the density profiles
    and \citet{delpopolo_01} and \citet{wechsler_02} the evolution of
    the concentration parameter.

  \subsubsection{The dark matter halo formation history}
 
\begin{figure}
      \resizebox{\hsize}{!}{\includegraphics{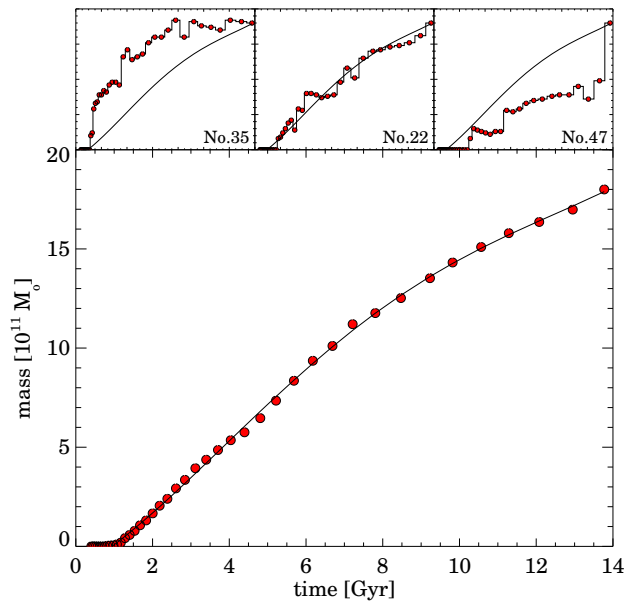}}
      \caption{The dots shows the average mass from 96 different halo
               merging trees, which all result in a $1.8 \times
               10^{12} \msun$ halo. The smooth line shows the
               approximation which is used in our numerical model. The
               most extreme halo merging trees and that which shows
               the smallest deviations from the average are shown in
               the smaller upper panels. Data from the
               VIRGO-GIF-project \citep{kauffmann_99}}
      \label{MS2866f1}
    \end{figure}

    In the hierarchical formation scenario, a dark halo may form early
    and rapidly or grow slowly with time. For the model described in
    this paper we take an average halo formation history, because at
    present we cannot simulate the evolution of a galaxy for a large
    number of different halo formation histories. We choose a \LCDM
    universe with $\Lambda = 0.7$, $\Omega = 0.3$, a Hubble constant
    $h_0 = 0.7$ and a baryonic-to-dark matter ratio of 1:5. From the
    cosmological N-body simulations of the VIRGO-GIF-project
    \citep{kauffmann_99}, we extracted the merging trees of 96 halos,
    all with a final mass of $1.8 \times 10^{12} \msun$. Figure
    \ref{MS2866f1} shows the average halo mass growth, which is in
    agreement with the universal mass accretion histories found by
    \citet{vandenbosch_02} and \citet{wechsler_02}. The individual
    merging histories can be quite different; two extreme cases with
    early and late mass growth are shown in the small panels at the
    top left and right of Fig.~\ref{MS2866f1}.
 
    We assume that the growing dark halo has the universal density
    profile as proposed by \citet{navarro_97}, and that its mass is
    accreted in spherical shells. Given the current halo mass and the
    evolution of the critical density, we can determine the outer halo
    radius $r_{200}$ \citep[see][]{mo_98}. A free parameter that
    remains is the mass concentration $c$, which in the present
    simulation is assumed to be $c = 9.2$. At redshift $z = 0$, this
    leads to a circular velocity of $v_{\mathrm{circ}} = 179 \kms$ at
    the $r_{200} = 250 \kpc$ radius. Inside a galactocentric radius of
    $8.5 \kpc$, the final mass of dark matter is $4 \times 10^{10}
    \msun$, comparable to the upper limit of the dark matter content
    inside the solar circle \citep[$5.2 \times 10^{10}
    \msun$;][]{navarro_00a}.
  
    Treating the growing dark matter halo as a spherical background
    potential that does not respond to the baryonic component
    minimizes gravitational torques during the collapse and thereby
    circumvents the well-known angular momentum problem
    \citep{navarro_00b}.  The accretion in this model is smoother as
    compared to hierarchical models, where a fraction of the mass
    accretion takes place through merging dark matter fragments,
    unless the gas in these hierarchical fragments is dispersed by
    early star formation and feedback.

  \subsubsection{The angular momentum of the dark matter}

    During the hierarchical clustering, tidal torques induce shear
    flows which lead to rotating proto-galactic systems. N-body
    simulations show that, independent of the initial perturbation
    spectrum and the cosmological model, the spin parameter of the
    dark halos is in the range $\lambda = L |E|^{0.5} G^{-1} M^{-2.5}
    = 0.02 - 0.11$ with an average value of $\lambda = 0.04 - 0.05$
    \citep{barnes_87, steinmetz_95b, cole_96, vandenbosch_98,
    gardner_01}.

    \citet{bullock_01} studied the angular momentum profiles of dark
    halos in a \LCDM cosmology. They found that the spatial
    distribution of the angular momentum in most of the halos show a
    certain degree of cylindrical symmetry, but that a spherically
    symmetric angular momentum distribution with a power-law $j(r)
    \propto r^{1.1 \pm 0.3}$ is also a good approximation. We consider
    the following two rotation fields for the halo:
    \begin{equation}
      v_{rot}^{halo} = v_0 { \sqrt{x^2+y^2} \over \sqrt{x^2+y^2+z^2} +
      d_0} \equiv v_0 {|r| \over |d| + d_0} \label{equlab_01}
    \end{equation}
    \begin{equation}
      v_{rot}^{halo} = v_0 { \sqrt{x^2+y^2} \over \sqrt{x^2+y^2} +
      d_0} \equiv v_0 { |r| \over |r| + d_0}
      \label{equlab_02}
    \end{equation}
    $v_0$ and $d_0$ are constants which determine the total amount of
    angular momentum and the rotation curve near the rotation axis,
    respectively.
 
\begin{figure}
      \resizebox{\hsize}{!}{\includegraphics{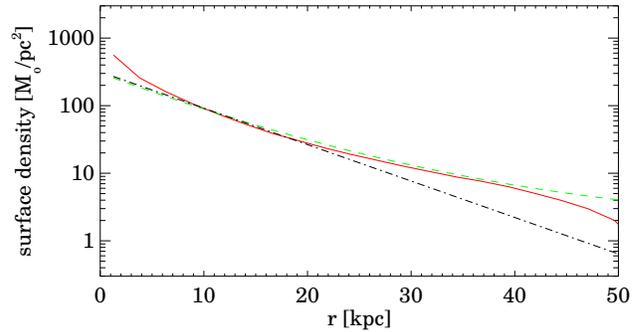}}
      \caption{Surface density of the galactic disk if the specific
               angular momentum is conserved during the proto-galactic
               collapse in our NFW halo model.  The cylindrical
               rotation leads to an exponential disk in the inner $20
               \kpc$ (dashed curve), while the spherical rotation
               produces an exponential disk with an additional mass at
               the centre (full line). The thin dash-dotted line shows
               an exponential disk with a scale length of $8 \kpc$.}
      \label{MS2866f2}
    \end{figure}

    Assuming that initially the mass and velocity distributions of the
    baryonic matter coincide with those of the dark matter, and that
    baryonic mass and specific angular momentum is conserved during
    the collapse to a disk \citep{mestel_63, fall_80}, we can estimate
    the disk surface density profiles corresponding to these rotation
    fields. Figure \ref{MS2866f2} shows the results, calculated for
    collapse in the pure NFW potential.  As already pointed out by
    \citet{fall_80}, disks with exponential surface density profiles
    are found when the halo has a rising rotation curve at small radii
    and a nearly constant rotation velocity in the outer parts. In the
    present case, the cylindrical halo rotation model
    (eqn.~\ref{equlab_02}) also produces an exponential disk, at least
    in the inner $20 \kpc$.  The spherical rotation model
    (eqn.\ref{equlab_01}) instead leads to higher surface densities
    near the galactic centre which favours the formation of a
    bulge. In the following we use the spherical rotation model with a
    spin parameter $\lambda = 0.05$, to fix the initial angular
    momentum distribution of the baryonic matter.

  \subsection{The Stars}

    The stellar motions have a considerable influence on the galactic
    evolution, since the time-delayed stellar mass loss and energy
    release connect different regions and also different galactic
    evolution phases. The approximately $10^{11}$ stars in a large
    galaxy form a collision-less system. Each star moves on its own
    orbit, which is determined by the conditions under which the star
    formed and the evolution of the galactic gravitational
    potential. We use a particle-mesh method \citep{hockney_88}, in
    which the passing of the stellar particles to the mesh and also
    the interpolation of the gravitational forces at the particle
    positions is done with a cloud-in-cell method. This method is fast
    and second order accurate, but it is limited by the spatial
    resolution of the underlying grid, which at present is $370 \pc$.

    Each stellar particle represents an ensemble of stars with a
    stellar mass function. In the present model, we assume a Salpeter
    initial mass function (IMF) with lower and upper mass limits of
    $0.8 \msun$ and $50 \msun$, and with an additional lock-up mass
    fraction of 60\%. The lock-up mass equals the mass of all stars
    below $0.8 \msun$. During a simulation the star formation rate is
    calculated on the spatial grid, and in timesteps of $10 \myr$
    representative samples of 500 new stellar particles are
    created. This yields 614\,500 particles at the end of the
    simulation. Depending on the star formation rate a single particle
    represents $10^4$ to $10^7$ stars. The initial chemical
    composition and velocity of a stellar particle is derived from the
    chemical composition, velocity and velocity dispersion of the
    parent molecular cloud medium.

    We assume that the evolution of the stars is determined only by
    their initial mass and metallicity, neglecting effects of stellar
    rotation or star-star interactions. From the stellar evolution
    models of \citet{maeder_89,schaller_92} and \citet{schaerer_93} we
    get the lifetimes, energy releases and mass losses on the main
    sequence. We can subdivide stars into three main classes.

    First, the low mass stars which stay on the main sequence during
    the whole galactic evolution. These stars become noticeable only
    by their gravitational forces and because they lock-up a major
    fraction of the galactic mass. The total lock-up mass has
    influence on the galactic gas content and therefore on the
    sequence of stellar generations. It is important in smaller
    galaxies where SN-driven winds reduce the mass in gas, but not the
    mass locked up in low mass stars.

    The second class of stars are intermediate in mass. When they
    leave the main sequence, they loose a substantial fraction of the
    initial mass (asymptotic giant branch and planetary nebula phase);
    however, the energy release of these stars is small compared to
    other energy sources.
 
    The third class of stars are the massive stars, typically with
    masses in excess of $8-10 \msun$. These stars are a major source
    of both energy and mass return in a galaxy. The UV-radiation and
    the kinetic energy release during the final explosion (SN type II)
    ionize and accelerate the ISM. This can trigger large-scale gas
    flows and star formation which both influence the galactic
    evolution.

    A rare but important galactic event is the explosion of SNe of
    type Ia. The most probable progenitor candidates are close binary
    stars consisting of a white dwarf and a main sequence star. We
    include this process, since it is a major source of iron peak
    elements \citep{nomoto_84}.

  \subsection{The Interstellar Medium}
  \label{ISM}

    The ISM in galaxies exists at different densities and
    temperatures, and it shows phase transitions on short timescales
    compared to the age of a galaxy. In the present galactic model we
    use a simplified ISM description based on the three-phase model of
    \citet{mckee_77}. The three phases are a cold, a warm and a hot
    phase, with typical temperatures of $80 \kelvin$, $8000 \kelvin$
    and $10^6 \kelvin$, respectively. The cold phase is found in the
    dense cores of molecular clouds. These cores are embedded in warm
    neutral or ionized gas envelopes. Outside the clouds the space is
    filled with hot, dilute gas.  The three-phase model is a
    simplification, but it is realistic enough to use it in a global
    galactic model. A critical discussion of the weak and strong
    points of the three-phase model can be found in \citet{mckee_90a}.

    Essentially, the temperatures and densities of the ISM phases
    determine most of the physical processes in the ISM. However, the
    geometry of the clouds is also important for processes that take
    place in the phase transition regions (e.g.~evaporation and
    condensation) or that depend on the sizes of the clouds explicitly
    (e.g.~cloud-cloud collisions and star formation). In our model, we
    assume spherical clouds which follow the mass-radius relation of
    \citet{elmegreen_89}
    \begin{equation}
      {M \over R^2} = 190 \left({\msun \over \pc^2}\right) \sqrt{P_4}.
      \label{equlab_03}
    \end{equation}
    $P_4 = \press$ is the pressure in the ambient ISM.  The mean cloud
    mass is $10^5 \msun$; this is derived from the cloud mass spectrum
    of \citet{dame_86} for star forming giant molecular clouds in the
    mass range of $10^4 \msun$ to $10^6 \msun$.

    In the present dynamical description, we use only a two-phase
    model for the ISM, containing hot gas with embedded (cold+warm)
    clouds. The motions of both components are described by the
    time-dependent hydrodynamical equations (mass, momentum and energy
    conservation). Similar as in the two-dimensional chemo-dynamical
    models \citep{samland_97} we use a fractional step method to split
    the problem into source, sink and transport steps. The problem of
    the three-dimensional transport is broken down into a number of
    one-dimensional problems for the densities, metallicities, momenta
    and energies (dimensional splitting). Using the van Leer advection
    scheme \citep{vanleer_77} combined with the consistent advection
    method \citep{norman_80} and the Strang splitting
    \citep{strang_68}, the advection is second order accurate. In the
    Strang splitting the three-dimensional transport is split into a
    sequence of five one-dimensional transport steps:
    \begin{equation}
      \Delta t \vec{A} = ({\Delta t \over2}\vec{A_x})({\Delta
      t\over2}\vec{A_y}) (\Delta t \vec{A_z})({\Delta
      t\over2}\vec{A_y})({\Delta t\over2}\vec{A_x})
      \label{equlab_04}
    \end{equation}
    where $\Delta t$ is the time step length, $\vec{A_x}$,
    $\vec{A_y}$, $\vec{A_z}$ are the differential operators which
    describe the transport in $x$, $y$, $z$ direction, and $\vec{A}$
    is the operator for the three-dimensional problem. Approximately
    the same accuracy can be achieved by a simpler splitting scheme
    which uses an alternating sequence of only three one-dimensional
    transport steps:
    \begin{eqnarray}
      (\Delta t \vec{A})(\Delta t \vec{A}) = \left[ (\Delta t
      \vec{A_x})(\Delta t \vec{A_y})(\Delta t \vec{A_z}) \right]
      \nonumber\\ \left[ (\Delta t \vec{A_z})(\Delta t
      \vec{A_y})(\Delta t \vec{A_x}) \right]
      \label{equlab_05}
    \end{eqnarray}
    For the $\sim 10^6$ transport steps in our galactic evolution
    simulations, both splitting schemes lead to identical
    results. However, the second method is faster by a factor of 1.6.

  \subsection{The interaction network}
 
    The interactions between stars and ISM are very important for the
    galactic evolution. They determine the evolution timescale of a
    galaxy to a large extent. In the following subsections we give a
    short overview about the processes which we take into account, and
    discuss the influence of the free parameters in the
    parametrization of these processes.

  \subsubsection{Star formation}
 
    Star formation is obviously one of the most important processes
    during the galactic evolution, but it is one of the least
    understood. Early on, \citet{schmidt_59} found that the star
    formation rate in the solar neighbourhood varies with the square
    of the surface gas density $\Sigma_\mathrm{gas}$ of the ISM. More
    recently, \citet{kennicutt_98} found a star formation rate for
    disk and starburst galaxies which is proportional to
    $\Sigma_\mathrm{gas}^{1.4 \pm 0.15}$, with a sharp decline below a
    critical surface density threshold ($\approx 3$-$13
    \msunpcz$). Remarkably, this simple law describes star formation
    in galaxies for which the average gas consumption time can vary
    from $3 \cdot 10^8 \yr$ (starbursts) to $2 \cdot 10^9 \yr$ (normal
    disks).

    Motivated by these findings we use the following star formation
    law:
    \begin{equation}
      \SFR = {\rhcld \over \tsf} = {\rhcld \over \csf \tdyn}
      \label{equlab_06}
    \end{equation}
    Here $\rhcld$ is the average density of the cloudy medium, $\tsf$
    the star formation timescale, and $\tdyn$ the dynamical timescale
    for the average star forming cloud mass. For the cloud model
    described in Section \ref{ISM} the dynamical timescale of the
    average star forming cloud is
    \begin{equation}
      \tdyn = \sqrt{3 \pi \over 32 G \hat \rhcld} = 16.2 \cdot
      P_4^{-3/8} \label{equlab_07}.
    \end{equation}
    Molecular clouds do not collapse and form stars on a single
    dynamical timescale, because magnetic fields or turbulent motions
    stabilize the clouds.  To account for these processes, we include
    the factor $\csf$. If we take into account that a part of the
    clouds is only confined by pressure and that the timescales for
    ambipolar diffusion and dissipation of turbulent energy are much
    longer than the dynamical timescale, we expect $\csf \gg 1$.
    $\csf$ can be estimated from the observed number of OB stars in
    the solar neighbourhood \citep{reed_01}. Assuming that the stellar
    mass function follows a Salpeter law with lower and upper mass
    limits of $0.1 \msun$ and $50 \msun$, we derive a local star
    formation rate of $1.8 \times 10^{-5} \msunpcdmyr$. From the local
    ISM density and pressure ($\rho \approx 0.035 \msunpcd \approx 1
    \cmd$, $P_4 \approx 1$), we find $\csf = 120$, which means that
    the on average the star formation timescale exceeds the dynamical
    timescale by factor of 120. The final star formation law can then
    be written as
    \begin{equation}
      \SFR = \etasf \rhcld P_4^{3/8}
      \label{equlab_08}
    \end{equation}
    with a star formation (gas depletion) timescale $\tsf \approx 1 /
    \etasf = 1.9 \gyr$ in the local Galactic disk.  The star formation
    rate is proportional to $\rhcld^{1.375}$ if the cloudy medium is
    in pressure equilibrium ($P_4 \propto \rhcld$). This is the
    quiescent star formation mode in the model. During the collapse or
    if large fragments merge, the pressure can increase and the star
    formation can be more efficient.

\begin{figure}
      \resizebox{\hsize}{!}{\includegraphics{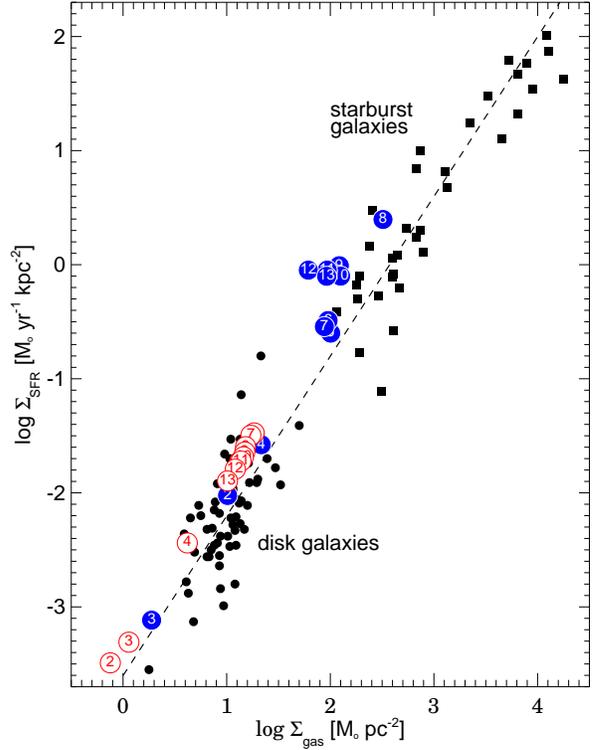}}
      \caption{Star formation rate in the forming disk galaxy model
               described in Section \ref{history}, versus the
               projected surface density of the cloudy medium in a $20
               \kpc$ sphere. Open numbered circles are for average
               star formation rates and surface densities within $20
               \kpc$; the number denotes the galactic age in Gyr. The
               filled numbered circles are for the most prominent star
               formation region at the same galactic ages. The black
               dots and black squares denote the data from
               \citet{kennicutt_98} for disk galaxies and starburst
               galaxies. The dashed line is Kennicutt's mean
               relation.}
      \label{MS2866f3}
    \end{figure}

    Figure \ref{MS2866f3} shows the resulting star formation rate as a
    function of the surface density of gas, in the forming disk galaxy
    model described in Section \ref{history}.  Both the average star
    formation rates and the star formation rates in the strongest star
    formation regions are slightly higher than, but consistent with
    \citeauthor{kennicutt_98}'s \citeyear{kennicutt_98} data and mean
    relation, over a range of surface densities extending from normal
    disk galaxies into the starburst galaxy region. Had we calibrated
    the factor $\etasf$ on Kennicutt's diagram, the resulting value
    would have been about a factor of 2 smaller than that based on the
    solar neighbourhood OB stars. The typical scatter in Kennicutt's
    plot corresponds to a factor of 2.2 dispersion in $\etasf$.

  \subsubsection{Massive stars and supernovae type II}

    Massive stars and SNe influence the galactic evolution in several
    ways. They heat the ISM, produce shells and filaments and stir up
    the cloudy medium. In addition, they are the most important source
    of heavy elements and they determine the chemical evolution of a
    galaxy. The lifetimes of the massive stars range from 3 to $30
    \myr$ before they finally explode as SNe II. Only a stellar
    remnant of $\approx 2 \msun$ remains. The rest of the mass is
    blown away by stellar winds or expelled during the explosion.  On
    average a massive star on the main sequence emits about $6 \times
    10^{51} \erg$ in photons \citep{schaerer_98} which ionize the
    surrounding ISM and regulate the star formation. During the SN
    explosion the star releases another $10^{51} \erg$, creating a
    fast expanding bubble. This influences the star formation process,
    since it decreases the gas density in the surroundings of the SN
    and it can trigger star formation in other regions.

    In the model we can use only a simple description of these
    processes.

    (i) We assume that the mass return of a massive star takes place
    instantaneously at the end of its evolution, when the SNe
    explodes. This is a good approximation, because the time between
    the stellar mass loss by a wind and the ejection of the SN shell
    is short compared to galactic timescales.

    (ii) We assume that the UV photons emitted by the massive stars
    are absorbed and reradiated by the dense cloud medium. Since the
    cooling time of this gas is short compared to the lifetime of the
    massive stars (the typical timescale which we resolve in our
    models), we do not simulate this heating and cooling process, but
    assume that the clouds are in thermal equilibrium.

    (iii) The SN explosion energy ($10^{51} \erg$) is released in a
    short time and produces a bubble of hot gas and an expanding
    shell. The shell interacts with the clouds and by this increases
    the velocity dispersion $\sicld$ of the cloudy medium.  The SN
    energy is released locally, and we assume that $\etaicm = 95\%$ of
    the SN energy heats the intercloud gas, and that the remaining
    $\etacld = 5\%$ goes into the velocity dispersion of the cloudy
    medium \citep{mckee_77}. Thus
    \begin{equation}
      \dot \ekicm = \etaicm \cdot \dot n_{\mathrm sn} \cdot 10^{51}
      \erg \approx \etaicm \cdot \SFR \cdot \esn,
      \label{equlab_9}
    \end{equation}
    \begin{equation}
      \dot \ekcld = \etacld \cdot \dot n_{\mathrm sn} \cdot 10^{51}
      \erg \approx \etacld \cdot \SFR \cdot \esn.
      \label{equlab_10}
    \end{equation}
    In the second half of these equations we have related the heating
    rates to the star formation rate; here $\esn = 2.7\cdot10^5
    \pczmyrz$ is the specific SN energy (per mass of stars formed for
    one supernova II to occur).  The parameters $\etaicm$, $\etacld$
    and $\esn$ are not free, but are fixed in a narrow range by
    observations and supernova models.

    (iv) SNe are the most important sources of heavy elements. The
    lack of self-consistent explosion models of massive stars causes
    uncertainties in chemical yields by factors of two and more.
    However, the detailed chemical composition is not important for
    the galactic evolution.  We therefore use a fiducial chemical
    element which traces the fraction of heavy elements produced in
    the hydrostatic burning phases and during the SN explosion of
    massive stars. With the yield tables of \citet{woosley_95} and
    \citet{thielemann_96}, or the yield approximations of
    \citet{samland_98} we can convert the abundances of the fiducial
    SN II element into real chemical abundances. This method has the
    advantage that it is not necessary redo the dynamical simulation
    when new yield tables become available.

  \subsubsection{Supernovae type Ia}

    The progenitors of Type Ia SN are believed to be close binary
    stars consisting of an intermediate mass star and a white
    dwarf. In the present model only a small fraction of the $1.5
    \msun$ - $8.0 \msun$ stars explode as SNe Ia. This fraction is
    determined by the number ratio of SNe type Ia to type II. We take
    a ratio of 1:8.5 \citep{samland_98} which is consistent with the
    observed SN rates in galaxies \citep{tammann_94} and can explain
    the observed iron abundances and the Galactic evolution of the
    [$\alpha$/Fe] ratio. In our model both the energy released by a SN
    Ia ($10^{51} \erg$) and the mass of the white dwarf progenitor
    ($0.6 \msun$) are given to the hot ISM. Because the mass return of
    SN Ia is small compared to other stars and the total energy
    released is an order of magnitude smaller than that from SNe type
    II, the SNe Ia do not influence the dynamical evolution
    significantly. However, SNe type Ia are important for the chemical
    evolution, because they are a main source of iron-peak, r- and
    s-process elements. In the same way as for the SN II, we include a
    fiducial chemical element to trace the enrichment with SN Ia
    nucleosynthesis products. For the conversion into real abundances
    we use the chemical yield table of \citet{nomoto_84}.

  \subsubsection{Intermediate mass stars}

    Intermediate mass stars act as a mass storage for the ISM. The
    mass loss of these stars is significant only at the end of the
    stellar evolution, when these stars enter the AGB and planetary
    nebulae phase. Intermediate mass stars end as white dwarfs with
    masses between $0.5 \msun$ and $1 \msun$. Unlike the massive
    stars, the intermediate mass stars eject only weakly enriched gas
    \citep{vandenhoek_97} and, in total, they return twice as much
    mass to the ISM as the massive stars, however with a much lower
    energy and a significant time delay. In the model we neglect the
    radiation of the intermediate mass stars, because they do not heat
    the ISM efficiently, but include the mass return and the
    enrichment in terms of a third fiducial element.

  \subsubsection{Radiative cooling of the hot gas}

    The heating of the ISM by SNe is mainly balanced by radiative
    energy losses. This cooling process determines the temperature and
    pressure of the hot ISM and by this has influence on the dynamics
    of the ISM. We use the metallicity and temperature dependent
    cooling functions $\lcool$ of \citet{dalgarno_72} and
    \citet{sutherland_93}. These cooling functions provide lower
    limits to the real energy loss, because they are calculated for a
    homogeneous gas in thermal equilibrium. In a real ISM with density
    fluctuations, the cooling rate can be enhanced by a factor
    $\etacool = 2.3-10$ \citep{mckee_77}, so that
    \begin{equation}
      \dot \ekicm = -\etacool \cdot \rhicm^2 \lcool.
      \label{equlab_11}
    \end{equation}
    We use an enhancement factor of $\etacool = 5$ to describe the
    radiative energy loss in the simulations. The radiative energy
    loss $\dot \ekicm$ scales with the square of the gas density
    $\rhicm$ because the ions are excited by collisions.

  \subsubsection{Dissipation in the cloudy medium}
  \label{dissipation}

    The cloudy medium is described as a hydrodynamic fluid
    characterized by its density, momentum, and kinetic energy. The
    individual clouds are assumed to be at a thermal equilibrium
    temperature. The kinetic energy density of the cloudy medium has
    source and sink terms from the heating by SNe and dissipation by
    cloud-cloud collisions. The description of the cloud-cloud
    collisions is based on the inelastic cloud collision model of
    \citet{larson_69}. There, the energy loss of the cloudy medium
    depends on the effective cross section $\Sigma = \ccoll \cdot \pi
    R^2$, which differs from the geometrical cross section by the
    factor $\ccoll$. For a medium with density $\rhcld$ and velocity
    dispersion $\sicld$ consisting of clouds of mass $M$ and radius
    $R$, the kinetic energy density then changes according to
    \begin{equation}
      \dot \ekcld = - \ccoll {8 \sqrt{\pi} \over 3} {R^2 \over M}
      \cdot \rhcld^2 \sicld^3.
      \label{equlab_12}
    \end{equation}
    With the cloud mass-radius relation of \citet{elmegreen_89} we
    obtain
    \begin{equation}
      \dot \ekcld = - \etacoll \cdot P_4^{-1/2} \cdot \rhcld^2 \sicld^3
      \label{equlab_13}
    \end{equation}
    with $\etacoll = 0.025$ for $\ccoll = 1$ (geometrical cross
    section). Unfortunately, $\etacoll$ is not well constrained,
    because magnetic fields, gravitational focusing, and self-gravity
    can influence the effective cloud cross section significantly.
    $\etacoll$ is the most uncertain parameter in our ISM description.

  \subsubsection{Evaporation}

    The different phases of the ISM exchange mass by evaporation as
    well as condensation and cloud formation processes. These
    processes depend mainly on the sizes of the cold clouds and the
    density and temperature of the hot ambient gas. As in the
    two-dimensional models \citep{samland_97}, we use the model of
    \citet{mckee_90} and \citet{begelman_90} for the description of
    the evaporation,
    \begin{equation}
      \rheva = 4.35\cdot 10^{-16} {R \over M} \rhcld \cdot \teicm^{5/2}.
      \label{equlab_14}
    \end{equation}
    For $10^5 \msun$ clouds in the simple spherical cloud model of
    Section \ref{ISM} we obtain an evaporation rate of
    \begin{equation}
      \rheva = \etaevap P_4^{-1/4} \rhcld \cdot \teicm^{5/2}
      \label{equlab_15}
    \end{equation}
    with $\etaevap = 1.0 \times 10^{-19}$. The value of $\etaevap$
    depends on the cloud model (cloud shapes, mass spectrum).

  \subsubsection{Condensation and cloud formation}

    In small clouds, heat transfer by conduction is more efficient
    than cooling by radiation \citep{mckee_90}. In large clouds (cloud
    radius exceeds the Field-length) the cooling dominates and the
    clouds can gain mass by condensation. The timescale for
    condensation is of the order of the cooling time of the ambient
    hot medium \citep{mckee_90}. In addition, clouds can form when the
    hot gas becomes thermally unstable. This cloud formation process
    is important in high density regions (e.g., shocks) or in regions
    with no or low star formation activity. Since it is a cooling
    instability, this process also has a timescale which is
    approximately the cooling time. We use the following
    parameterization for the condensation and cloud formation process:
    \begin{equation}
      \rhcon = \etacond \cdot {\rhicm \over \tcool}
      \label{equlab_16}
    \end{equation}
    It turns out that effectively $\etacond$ is not a free
    parameter. Reasonable values for $\etacond$ are in the range of
    1.0 (cloud formation time = cooling time) to 0.3. Tests showed
    that if $\etacond < 0.3$, the hot gas can cool to temperatures of
    less than $10^4 \kelvin$ before condensation and cloud formation
    are efficient. We use a value of $\etacond = 0.5$, which
    guarantees a stable hot gas phase.

  \subsubsection{Self-regulation}

    The interaction network describes a self-regulated system in which
    moderate changes of the parameters do not alter the results
    significantly \citep[see also][]{koeppen_95}. Star formation is
    one example for such a self-regulation. An increase of the
    efficiency $\etasf$ by a certain factor will not increase the star
    formation rate by the same amount, because the ISM will be heated
    by the stars, which in turn decreases the amount of available cold
    gas to form stars. Another example is the phase transition between
    hot and warm gas. A rise in the heating rate of the hot gas leads
    to an enhanced evaporation of clouds. This increases the density
    of the hot gas and enhances the cooling. The temperature of the
    gas increases only by a small amount. If on the other hand the
    heating rate decreases, condensation and cloud formation set
    in. As a result, the hot gas density and therefore the cooling
    efficiency drops and the gas temperature will stay nearly
    constant.

    Some interactions are self-regulated by dynamical processes. For
    example, a higher star formation efficiency delays the settling of
    the clouds in the galactic disk, so that the average density of
    the cloudy medium is lowered which in turn decreases the star
    formation rate again.  For the moment we neglect this (important)
    dynamical self-regulation, discussing now only the influence of
    the important model parameters on the velocity dispersion of the
    clouds $\sicld$, the hot gas temperature $\teicm$, and the mass
    ratio between gas and clouds. We do this by calculating
    equilibrium conditions first numerically and then in an analytical
    approximation.

\begin{figure}
      \resizebox{\hsize}{!}{\includegraphics{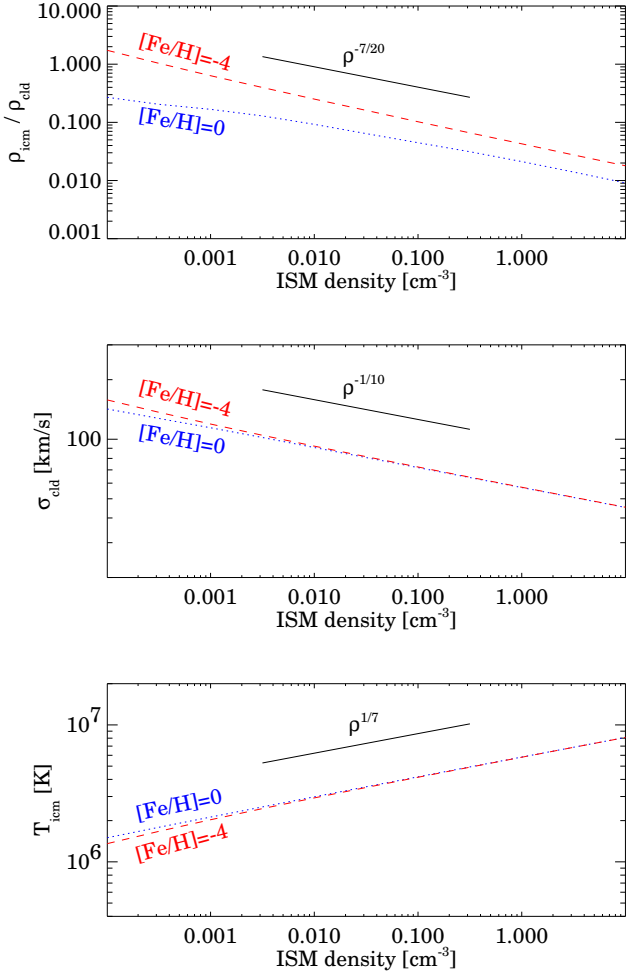}}

      \caption{The three panels show the density dependence of the
               intercloud to cloud mass ratio, the cloud velocity
               dispersion $\sicld$, and the hot gas temperature
               $\teicm$ in a closed-box model where dynamical effects
               are neglected. The dotted and dashed lines show the
               [Fe/H] = 0 (solar metallicity) and the [Fe/H] = -4
               cases and the full lines show the slopes given by the
               analytical approximations of
               eqns.~\ref{equlab_17}--\ref{equlab_19}.}
      \label{MS2866f4}
    \end{figure}

    The interaction network is a closed system of differential
    equations, which can be solved numerically to find equilibrium
    states for given densities of the ISM. The results are plotted in
    Fig.~\ref{MS2866f4}. As expected for a self-regulated system, we
    find that even density variations by five orders of magnitude have
    only small effects, changing $\sicld$ ($\teicm$) by factors of 3
    (7).

    For the analytical approach, we assume that, most of the ISM mass
    is in the cloudy medium and that the cloudy medium is in pressure
    equilibrium with the surrounding gas. Balancing the heating by SNe
    with the dissipation by cloud-cloud collisions and the cooling by
    radiation, and assuming that evaporation rates are equal to
    condensation rates, we get the following three relations:
    \begin{equation}
      \sicld \propto \left( {\etasf \etacld \over \etacoll}
      \right)^{4/5} \cdot \rho^{-1/10}
      \label{equlab_17}
    \end{equation}
    \begin{equation}
      \teicm \propto \left( {\etasf \etacld \over \etacoll}
      \right)^{2/7} \cdot \left( {\etasf \etacond \over \etaevap}
      \right)^{2/7} \cdot \rho^{1/7}
      \label{equlab_18}
    \end{equation}
    \begin{equation}
      {\rhicm \over \rhcld} \propto \left( {\etasf \etacld \over
      \etacoll} \right)^{3/10} \cdot \left({\etasf \over \etacool}
      \right)^{1/2} \cdot \rho^{-7/20}
      \label{equlab_19}
    \end{equation}
    Here $\rho$ is the average density of the ISM (clouds and hot
    gas). These equations confirm that there is only a weak dependence
    of $\sicld$ and $\teicm$ on the average ISM density. They are
    shown in Fig.~\ref{MS2866f4} for comparison with the numerical
    results.

    In the previous discussion, we neglected metallicity
    effects. However, metallicity can be important because it
    increases the cooling efficiency. The dashed lines in
    Fig.~\ref{MS2866f4} show the numerical equilibrium relations for an
    extremely metal-poor gas ([Fe/H] = -4). The effects of low gas
    metallicities are similar to a low cooling efficiency
    $\etacool$. In both cases, the gas temperature $\teicm$ stays
    nearly constant, because the lower cooling efficiency is balanced
    by a higher gas density $\rhicm$. The self-regulating character of
    the interaction network again is obvious. The equilibrium system
    is in a balance of forward and reverse reaction rates. Any stress
    that alters one of these rates makes the system shift, so that the
    two rates eventually equalize (Le Chatelier's principle).
  
    In spite of the self-regulating character of the interaction
    network it is necessary to solve the full interaction network
    during a simulation, because processes on different timescales are
    involved that compete with the dynamical evolution. For example,
    heating processes or cloud dissipation may take longer than
    dynamical changes. Equilibrium solutions of the interaction
    network can be used only to study the general behaviour of a
    star-gas system where dynamical processes are slow, but would not
    be able to describe, e.g., the outflow of hot gas.

  \subsubsection{Sensitivity to free parameters}

    Most of the parameters in this multi-phase ISM description
    ($\eta$'s) can be constrained either from theory or from
    observations. The two most important parameters in the model are
    the star formation efficiency $\etasf$, and $\etacoll$ which
    controls the dissipation rate. These parameters are of special
    interest because they both influence the velocity dispersion of
    the cloudy medium and thus the settling of the baryonic matter.

    In Table~\ref{tablab_01} we list the parameters of the interaction
    network together with their upper and lower limit values. These
    are discussed in the subsections above. For $\etaevap$ and
    $\etacoll$, where there is little constraint, we have used a large
    range to illustrate that even in this case the resulting changes
    to the equilibrium are not large.  It is obvious from the
    analytical model above that $\etasf$ ($\etacld$, $\etacoll$) have
    influence on $\sicld$, $\teicm$ and $\rhicm$, while $\etaevap$
    ($\etacond$) can change only $\teicm$, and by varying $\etacool$
    one can shift the gas-to-cloud mass ratio. Using the extreme
    values for the parameters given in Table~\ref{tablab_01} we obtain
    the uncertainties of $\sicld$, $\teicm$, and $\rhicm$ (column 5,
    6, and 7 of Table~\ref{tablab_01}). This shows that even large
    changes of the parameter values lead only to moderate variations
    in $\sicld$, $\teicm$, and $\rhicm$. From these numbers we
    conclude that $\sicld$, $\teicm$, and $\rhicm$/$\rhcld$ are
    accurate to a factor of 2 in our model.

    \begin{table*}
      \centering
      \begin{tabular}{|l|c|c|c|c|c|c|} \hline

        parameter & eqn. & range & std.\ value & $\sicld$ /
        $\sicld^{\rm std}$ & $\teicm$ / $\teicm^{\rm std}$ & $\rhicm$
        / $\rhicm^{\rm std}$ \\
        \hline\hline
        $\etasf$ & \ref{equlab_08} & $(5-30)\times 10^{-4}$ & $5.3
        \times 10^{-4}$ & 1.0-4.0 & 1.0-2.7 & 1.0-4.0 \\
        $\etacld$ & \ref{equlab_10} & 0.02 - 0.1 & 0.05 & 0.5-1.7 &
        0.8-1.2 & 0.8-1.2 \\
        $\etacool$ & \ref{equlab_11} & 2.3-10 & 5 & 1.0 & 1.0 &
        1.5-0.7 \\
        $\etacoll$ & \ref{equlab_13} & 0.013-0.5 & 0.025 & 1.7-0.1 &
        1.2-0.4 & 1.2-0.4 \\
        $\etaevap$ & \ref{equlab_15} & $(0.1-10) \times 10^{-19}$ &
        $10^{-19}$ & 1.0 & 1.9-0.5 & 1.0 \\
        $\etacond$ & \ref{equlab_16} & 0.3-1.0 & 0.5 & 1.0 & 0.9-1.2 &
        1.0 \\
        \hline
      \end{tabular}\\[1.5ex]
      \caption{Parameters of the model interaction network and the
               equations in which they are defined. Column 3 shows
               lower and upper limits for the parameters, column 4 the
               standard values used in the simulations, and columns 5,
               6, and 7 give the equilibrium $\sicld$, $\teicm$, and
               $\rhicm$ resulting with the maximum and minimum
               parameter values, normalized by those obtained for the
               standard model values.}
      \label{tablab_01} 
    \end{table*}

  \subsection{Gravitation}
 
    Gravitation is a long range force that couples the baryonic and
    the dark matter and it is the main driver of the galactic
    evolution. To determine the gravitational potential, we first map
    all stars, the different phases of the ISM and the dark matter
    onto one grid. Then, depending on the kind of grid (equidistant or
    logarithmically spaced) we solve the Poisson equation with a FFT
    (fast Fourier transformation) or a SOR (successive over-relaxation
    with Chebeyshev acceleration) method. The gravitational forces are
    calculated from the potential at every point.

  \subsection{Initial conditions and code set-up}
  
\begin{figure}
      \resizebox{\hsize}{!}{\includegraphics{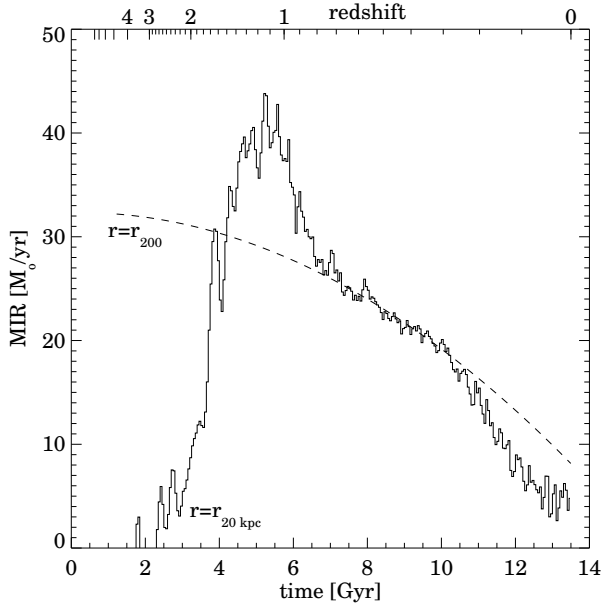}}
      \caption{Baryonic mass infall rate (MIR, full line) into inner
               $20 \kpc$ and across the $r_{200}$ radius (dashed
               line).}
      \label{MS2866f5}
    \end{figure}

    The present simulations used a logarithmic $81^3$ grid with a
    spatial resolution of $370 \pc$ in the inner parts and $5 \kpc$ in
    the outer halo. The timestep lengths, which in most cases are
    limited by the interaction processes, are of the order of
    $10^4-10^5 \yr$. The simulations start at a redshift $z = 4.85$
    corresponding to $t = 1.2 \gyr$ in a \LCDM cosmology. Initially
    the dark halo has a mass of $2.1 \times 10^{10} \msun$,
    distributed over a $30 \kpc$ sphere.  During the evolution, the
    halo grows by accretion to a size of $500 \kpc$ with an enclosed
    mass of $1.8 \times 10^{12} \msun$.  The baryonic-to-dark matter
    ratio of the accreted mass is fixed at 1:5, which finally amounts
    to $3 \times 10^{11} \msun$ of baryonic matter inside the
    $r_{200}$ radius. The accretion rate, taken from cosmological
    simulations \citep[][see Section \ref{model} and
    Fig.\ref{MS2866f5}]{kauffmann_99} decreases from $32 \msunyr$ at $z
    = 4.85$ ($1.2 \gyr$) to $8.1 \msunyr$ at $z = 0$ ($13.5
    \gyr$). The baryonic matter outside the $r_{200}$ radius consists
    of ionized ($T > 10^4 K$) primordial gas at the virial
    temperature. The cooling time of this gas is short compared to the
    collapse time. On this time-scale clouds form which can then
    dissipate kinetic energy and collapse inside the dark halo.
    Thereafter the collapse is only delayed by stellar feedback
    processes. We have confirmed that starting with a cloudy gas
    medium leads to a very similar evolution.  Initially, the accreted
    baryonic matter, like the dark matter, is in spherical rotation
    (eqn.\ref{equlab_01}) with a spin parameter $\lambda = 0.05$. It
    grows from inside-out and the mass with the highest specific
    angular momentum is accreted at late times.

  \section{The formation history}
  \label{history}
 
    The present simulation describes the formation and subsequent
    evolution of a disk galaxy in a slowly growing dark halo. The halo
    grows by accreting dark and baryonic matter in spherical shells
    with an accretion rate that is derived from cosmological
    simulations in a \LCDM universe \citep[see Section
    \ref{darkhalo}]{kauffmann_99}, with a baryonic mass fraction of
    1/6. The dark and baryonic matter have initial $\lambda = 0.05$
    and specific angular momentum distribution similar to that from
    \citet{bullock_01}.  Star formation, feedback, and gas physics of
    the baryonic component are described within a two-phase ISM model
    (see Section \ref{ISM}). The dynamics of the gas and stars is
    treated separately. The simulations start at $z = 4.85$ ($t = 1.2
    \gyr$) with a mixture of dark matter and gas, but no stars.  At
    redshift $z = 4.85$ the dark halo mass exceeds $2.0 \times 10^{10}
    \msun$, which is twice the mass resolution of the cosmological
    simulations on which our halo model is based.
 
  \subsection{Collapse and star formation history}

    The accreted primordial gas quickly reaches an equilibrium between
    the cold and hot phases with most of the mass in the cold clouds.
    It can then dissipate its kinetic energy and begins to collapse
    inside the dark halo. The strong collapse takes place in a
    redshift interval between $z = 1.8$ ($3.6 \gyr$) and $z = 0.8$
    ($6.6 \gyr$).  With a duration of $\sim 3 \gyr$, its period
    exceeds the free-fall time at the $r_{200}$ radius by a factor of
    3.4. This late and delayed collapse, which is in contradiction to
    simple collapse scenarios, is caused by stellar feedback in
    conjunction with the initially shallow gravitational potential of
    the halo.
 
    The baryonic mass infall rate (MIR) across the $r_{200}$ radius
    and into the inner $20 \kpc$ is shown in Figure~\ref{MS2866f5}. At
    early times the gas fallen through $r_{200}$ has not yet reached
    the inner $20 \kpc$ because of the delay from feedback. With the
    growing mass of baryons the dissipation increases and the infall
    into the centre accelerates, until the combination of feedback
    from the growing star formation rate and decreasing infall rate
    through $r_{200}$ reverses this trend and the MIR into $20 \kpc$
    begins to decrease again at $z \simeq 1.2$. At late times ($z \lta
    0.3$) a significant fraction of the infalling gas has too much
    angular momentum to arrive in the central $20 \kpc$, so that the
    rate of infall into $20 \kpc$ becomes lower than that through
    $r_{200}$.  The mass of the galaxy increases rapidly and at $z =
    1.2$ ($5 \gyr$) already $10^{11} \msun$ of baryonic matter has
    been accreted. At $z = 0$, 2/3 of the total baryonic mass of $3
    \times 10^{11} \msun$ has collapsed into the inner $20 \kpc$ of
    the dark matter halo and has formed the luminous galaxy, leading
    to a dark-to-baryonic matter ratio of 3\%, 10\% and 30\% inside
    galactocentric radii of 1, 3, and $10 \kpc$, respectively.
 
\begin{figure}
      \resizebox{\hsize}{!}{\includegraphics{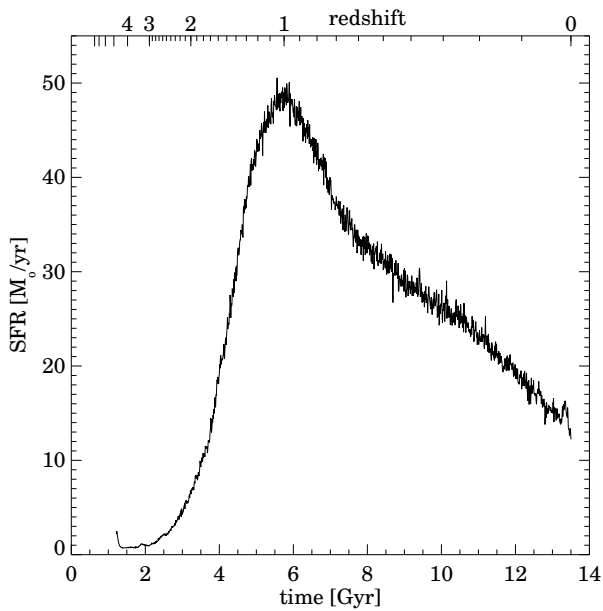}}
      \caption{Star formation rate (SFR) in the inner $20 \kpc$ of the
               dark halo. The star formation rate peaks slightly later
               than the MIR (Fig.~\ref{MS2866f5}). }
      \label{MS2866f6}
    \end{figure}

    The star formation rate integrated over the central $20 \kpc$ is
    shown in Figure~\ref{MS2866f6}. This increases steeply to a maximum
    of $\approx 50 \msunyr$ at $z = 1$ ($5.75 \gyr$). At that time,
    the gas consumption by the star formation process is balanced to
    80\% by infall and to 20\% by the stellar mass return. Afterwards
    the infall rate decreases to $5 \msunyr$, at $z = 0$, while the
    stellar mass return stays nearly constant at a rate of $10
    \msunyr$. This explains the different shapes of the curves in
    Figs.~\ref{MS2866f5},~\ref{MS2866f6} at late times.

  \subsection{Sequential formation of halo, bulge and disk}
 
\begin{figure*}[!t]
      \resizebox{\hsize}{!}{\includegraphics{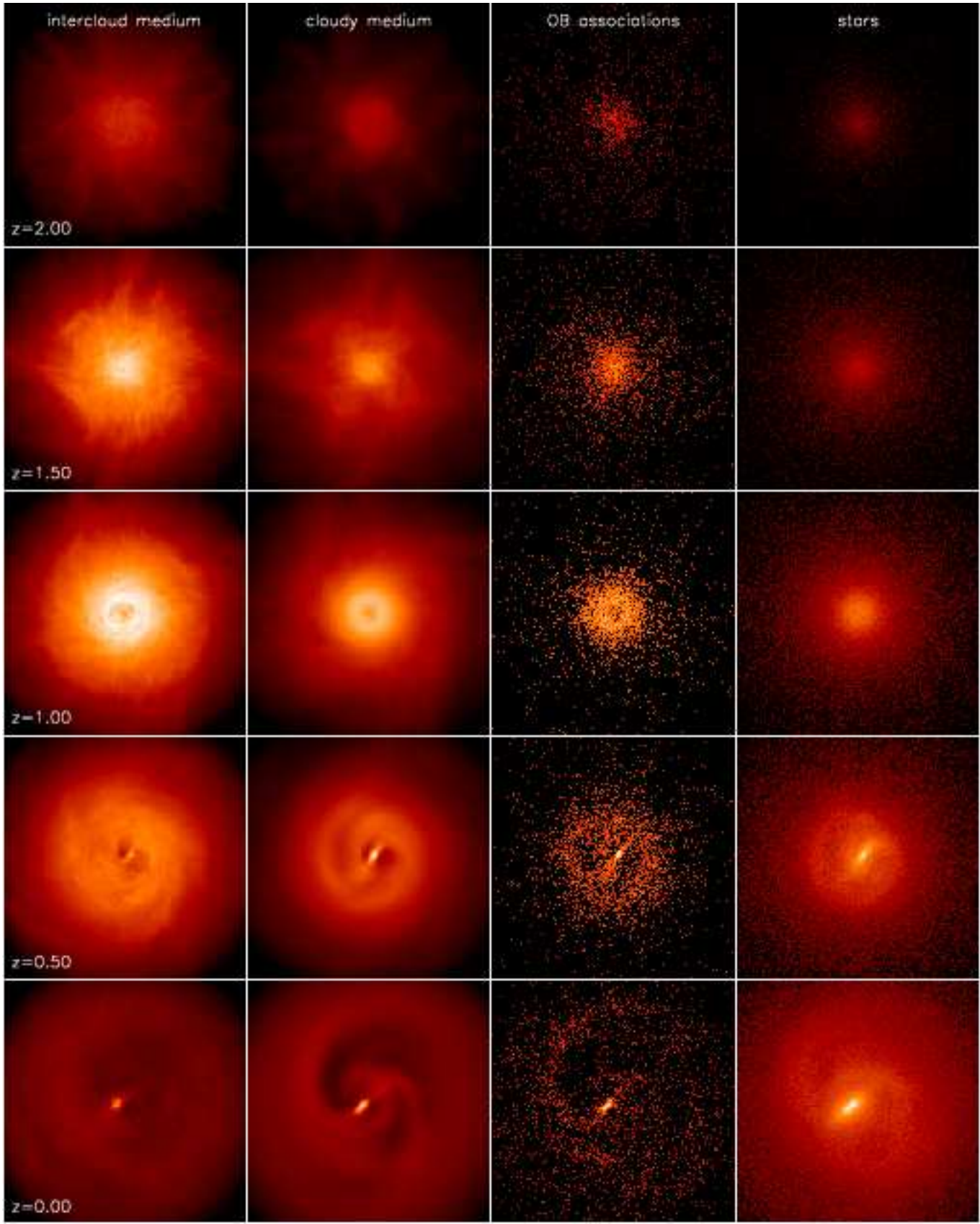}}
      \caption{Face-on surface density of the ionized gas, the cloudy
               medium, the OB-associations and the stars at different
               redshifts. Each column shows the evolution of one
               component between redshift $z = 2$ and $z = 0$. Each
               panel has a size of $50 \times 50 \kpc$.}
      \label{MS2866f7}
    \end{figure*}

\begin{figure*}[!t]
      \resizebox{\hsize}{!}{\includegraphics{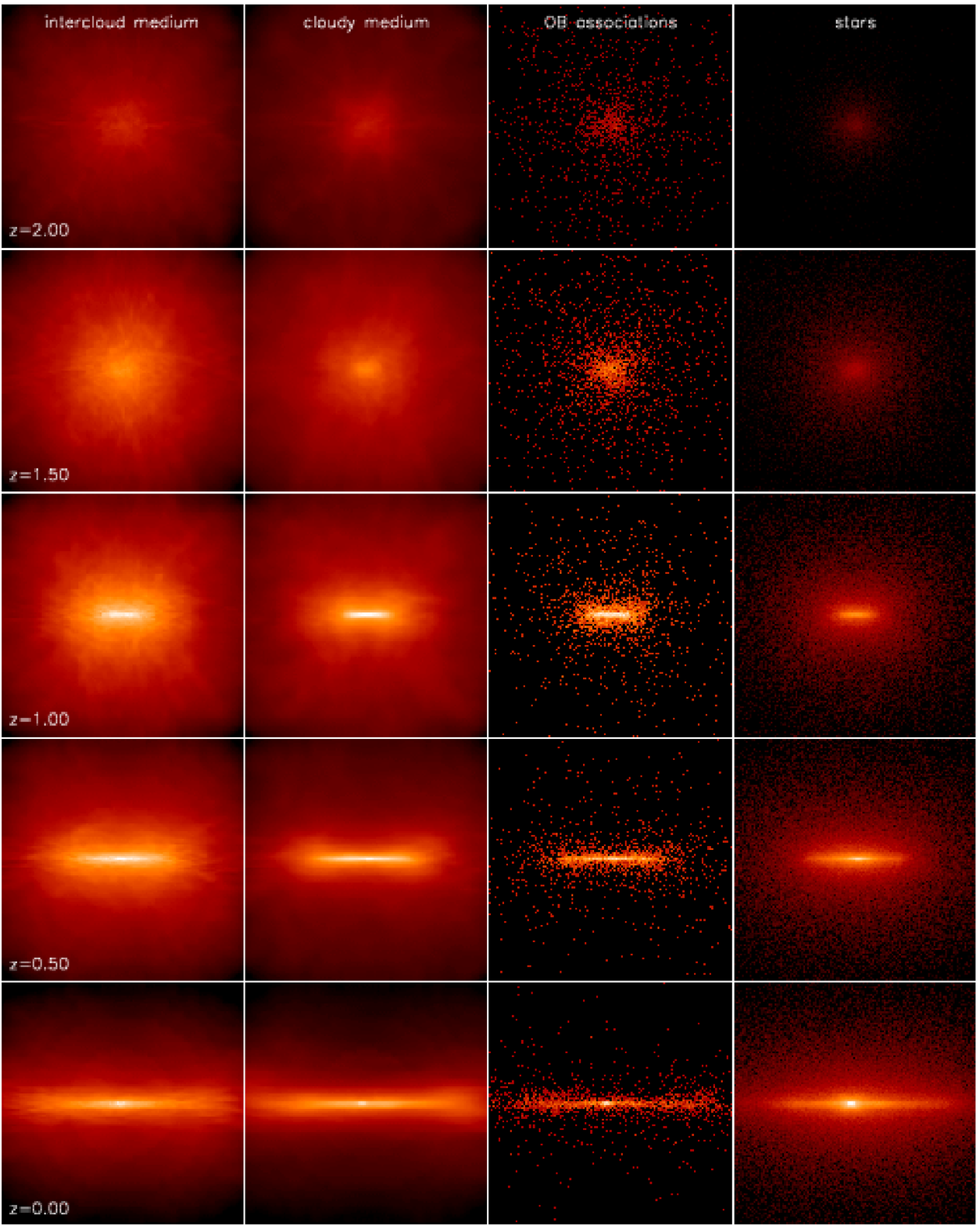}}
      \caption{Same as Fig.~\ref{MS2866f7}, but for the edge-on
      view. \smallskip \quad }
      \label{MS2866f8}
    \end{figure*}

    Figures~\ref{MS2866f7},~\ref{MS2866f8} show the spatial
    distributions of the gas and stars as a function of redshift.  In
    general terms, star formation out of the dissipating cloud medium
    proceeds from halo to disk and from inside out. Thus the earliest
    stars at redshifts $z > 2$ form in the whole volume limited by the
    $r_{200}$ radius (halo formation), with a concentration to the
    central bulge region. At a redshift of $z = 1.3$ ($4.7 \gyr$), a
    (thick) disk component first appears. Later the star formation is
    concentrated to the equatorial plane, and at late times most of
    the star formation occurs in the outer disk.

\begin{figure}
      \resizebox{\hsize}{!}{\includegraphics{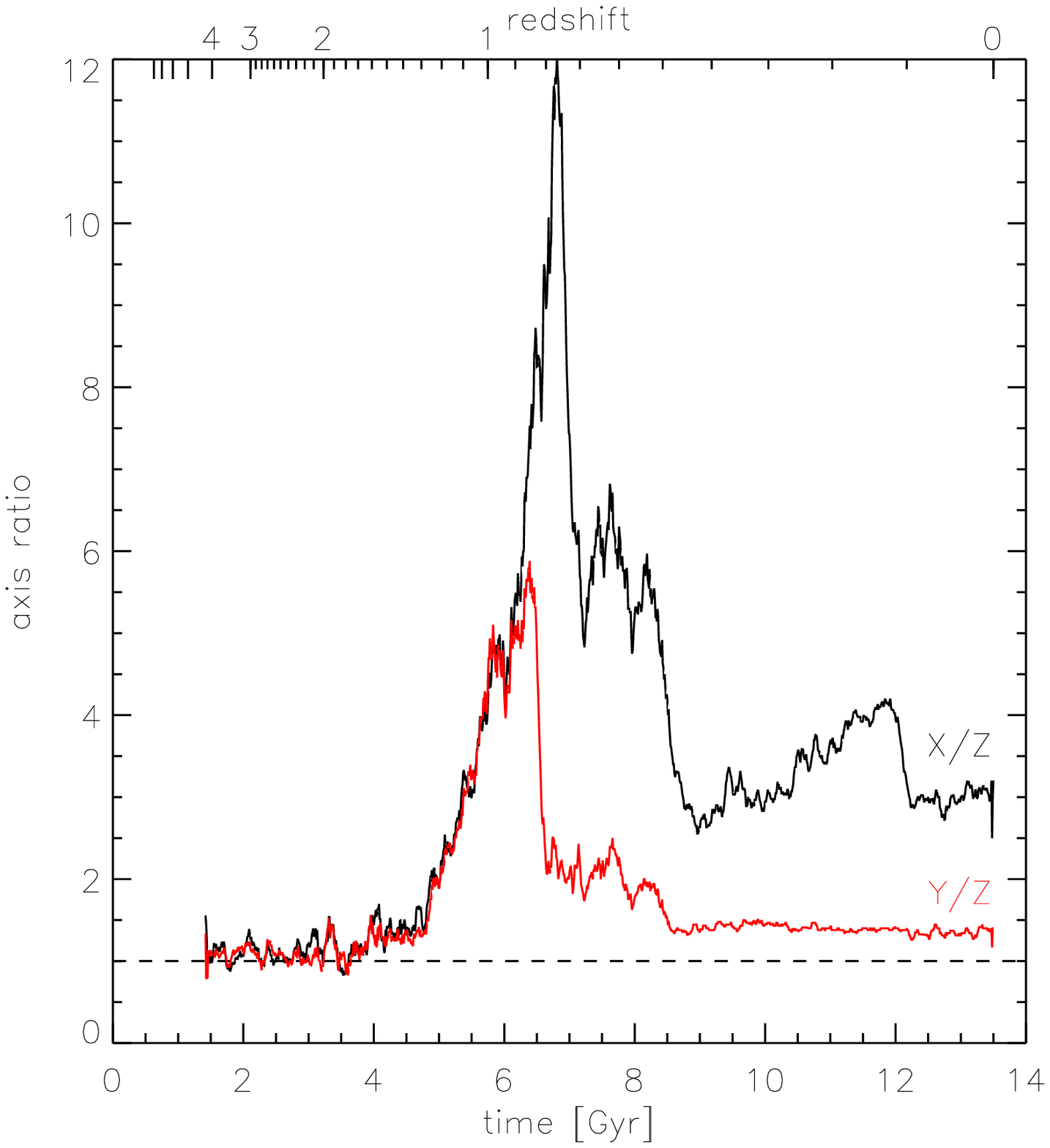}}
      \caption{Evolution of the bar-bulge axis ratios. z indicates the
               vertical direction and x and y the major and minor axes
               of the bar. The bar forms from a disk instability at
               time $6.4 \gyr$ and settles to a triaxial bulge with
               axis ratios $\sim$3:1.4:1 at late times.}
      \label{MS2866f9}
    \end{figure}

    Around $z = 1.2$ the infall of gas can no longer compensate the
    gas consumption by star formation in the inner disk. This is
    because the infalling molecular cloud medium now has higher
    specific angular momentum, so that its infall timescale becomes
    longer than the central star formation timescale. As a result, the
    region of highest star formation density moves radially outwards
    from the centre and a ring (Fig.~\ref{MS2866f7}, $z = 1$ panels)
    forms. This ring grows to a radius of $\sim 4 \kpc$ before, at
    redshift $z = 0.85$ ($6.4 \gyr$), the disk becomes
    unstable. Within $150 \myr$ the ring then fragments, and for a
    short time ($\approx 150 \myr$) a very elongated bar is formed
    with exponential major-axis scale-length of $5.9 \kpc$ and axis
    ratio of 12:2:1. Fig.~\ref{MS2866f9} shows that after this
    transient period the bar length decreases, and a bar-bulge with
    axial ratios $\sim$3:1.4:1 develops which includes the old bulge
    component formed in the early collapse.

\begin{figure}
      \resizebox{\hsize}{!}{\includegraphics{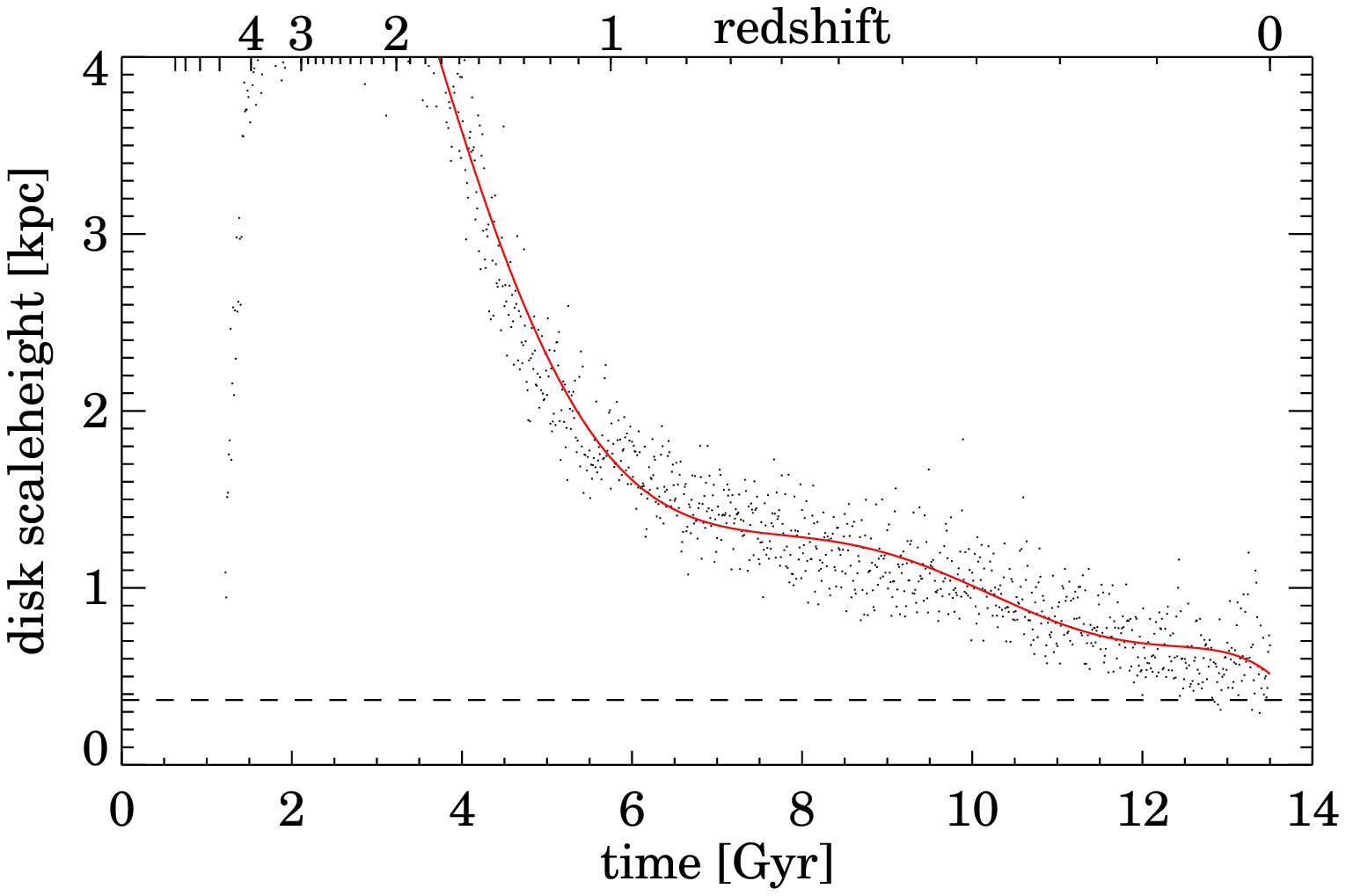}}
      \caption{Evolution of the vertical scale height of the stellar
               disk between radial distances $3-10 \kpc$. This part of
               the disk forms later than the average of all stars
               shown in Fig.~\ref{MS2866f5}.  The scale heights in
               small rings are shown as dots and the radial average is
               drawn as a full line.  The dashed line indicates the
               spatial resolution limit of $370 \pc$.}
      \label{MS2866f10}
    \end{figure}

     Already during this bar formation process the galaxy starts to
    build up the outer disk. This disk is the youngest component in
    the model galaxy, even though the oldest disk stars are as old as
    the halo stars. The disk grows from inside-out, because the early
    accreted mass has low specific angular momentum (see the last
    columns of Fig.~\ref{MS2866f7} and Fig.~\ref{MS2866f8}). In
    parallel, the vertical scale-height in a fixed radial range
    decreases; this is shown in Fig.~\ref{MS2866f10} for the range $3-10
    \kpc$. The more pronounced settling of the cloudy medium to the
    equatorial plane at late times is due to (i) more efficient
    cooling because of higher metallicity, (ii) higher gas density
    and, thus, dissipation rate for gas near the angular momentum
    barrier in a deeper potential well.  The final disk thickness in
    Fig.~\ref{MS2866f10} of $\approx 500 \pc$ is set by the resolution
    the present model.
 
    After the bar-bulge has formed, two trailing spiral arms appear in
    the disk which are connected to the bar-bulge. At the beginning
    the two spiral arms are symmetric and of the same size. However,
    in the further evolution a persistent lopsidedness of the galaxy
    develops.  The resulting off-centre motion of the bulge-bar brings
    one arm nearer to the bulge-bar, and the more distant spiral arm
    becomes more prominent over a wider winding angle
    (Fig.~\ref{MS2866f7}).  The lopsidedness may be produced by the
    dark halo which is here described as an external spherically
    symmetric potential component, even though in the inner galaxy the
    baryonic matter dominates the gravitational potential. The effect
    that such perturbations can produce lopsided galaxies is well
    known \citep{levine_98,swaters_99}.
 
    We expect this evolutionary sequence to be more or less typical of
    any dissipative collapse in a growing dark matter halo, unless it
    is interrupted by substantial mergers. The following modelling
    uncertainties may further modify the evolution as described above.
    (i) The assumed dissipation parameter is uncertain. If in reality
    dissipation is significantly more efficient than in the model, and
    in addition stars form only above a certain density threshold
    \citep{kennicutt_98}, the gas would fall into the disk much more
    rapidly without significant star formation. In this case formation
    of stars in the halo and thick disk may be substantially
    suppressed.  Also, in this case the velocity dispersion of the
    cloudy medium in the disk would decrease.  Then the gaseous disk
    may fragment in local instabilities before a global disk
    instability sets in, so that more material would dissipate to the
    centre and contribute to the bulge.  (ii) The precise angular
    momentum distribution (as compared to the universal average
    distribution assumed) will influence the detailed inside-out
    formation of the disk. For example, the formation of a ring in the
    present model would likely be suppressed by increasing the amount
    of low angular momentum gas falling in at redshift $z \simeq
    1$. (iii) When small-scale initial fluctuations lead to the
    formation of sub-galactic fragments that merge before the main
    collapse, the mass of the bulge component would be substantially
    greater, and its dynamical structure determined by the merging
    process \citep{steinmetz_02}.

  \subsection{Global chemical evolution}
  \label{chemical}
 
\begin{figure}
      \resizebox{\hsize}{!}{\includegraphics{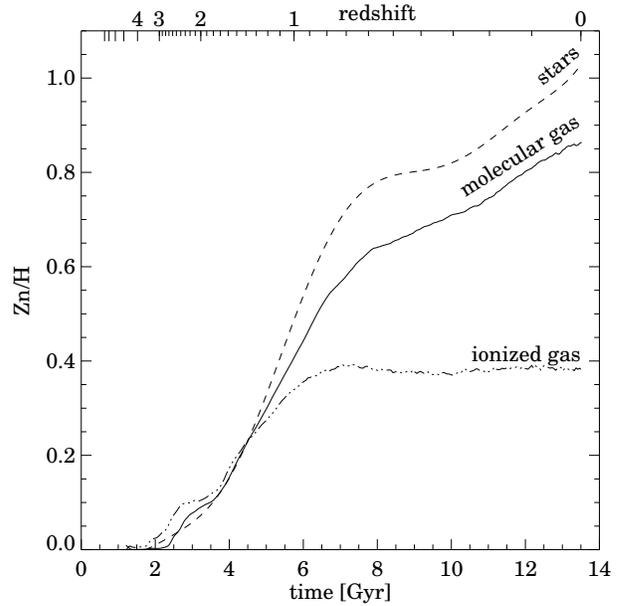}}
      \caption{Average \element{Zn}/\element{H} abundance (linear
               scale) of the ISM and the stars in the inner $20 \kpc$
               of the model galaxy.}
      \label{MS2866f11}
    \end{figure}

    The chemical evolution of the model proceeds in three
    steps. First, there is the collapse phase, lasting until $z \simeq
    0.6$.  Then two quasi-equilibrium phases follow in which first ($z
    = 0.6-0.3$) the mass infall and later ($z = 0.3-0.0$) the stellar
    mass return determines the star formation rate (see
    Figs.~\ref{MS2866f5}, \ref{MS2866f6}).  These three phases are
    clearly visible in the chemical enrichment history of the cloudy
    medium. This is shown in Fig.~\ref{MS2866f11}, where we have plotted
    the Zn/H metallicity normalized to the solar value because Zn is
    believed to not be depleted onto dust grains. The Zn/H abundance
    was reconstructed from the fiducial SN I and SN II elements as
    described in Section \ref{model}.
 
    The initial collapse of the baryonic material produces a first
    generation of stars which already at $z = 2$ has synthesized
    enough metals to increase the average metallicity to $ \approx
    -1$. During the late collapse phase the star formation rate
    exceeds the MIR and the metallicity increases fast. Then in the
    second phase, the star formation rate is proportional to the MIR
    and the metallicity increases only slowly. In the final phase, the
    star formation rate is again higher than the MIR and the
    metallicity increases more rapidly (Fig.~\ref{MS2866f11}).
 
    The chemical enrichment histories of the hot gas, cloudy medium,
    and stars (shown separately in Fig.~\ref{MS2866f11}) are very
    different from the predictions of multi-phase closed box models.
    In these models the hot ionized gas always has the highest
    metallicity, followed by the molecular cloud medium and the stars.
    This is different in our model, because stars, clouds and gas have
    different spatial distributions. Most of the ionized hot gas is
    located in the halo, while the molecular clouds and stars are
    preferentially found in the disk and bulge, where the gas
    densities and metallicities are high. After a brief initial period
    the average metallicity of the stars exceeds that of the cloudy
    medium, because the stars form from gas that is more concentrated
    to the equatorial plane and thus more metal-rich than the average
    cold gas medium.
 
    An interesting feature in Fig.~\ref{MS2866f11} is the constant
    [Zn/H] $\approx 0.4$ of the hot gas from redshift $z = 1$ until
    the present epoch. A large fraction of this gas occupies the
    galactic halo, where the star formation rate is low. Two processes
    keep the halo metallicity at a constant level: Gas flows from the
    bulge and the disk transport heavy elements into the halo. There
    it mixes with the existing ISM and with infalling low metallicity
    gas. In this way the metallicity in the halo gas can remain
    constant over a long time.
 
    A similar enrichment can be observed in the disk. During the
    collapse the star formation and the metal production is
    concentrated to the inner galaxy. From there, the metal-rich gas
    expands into the disk.  This disk pre-enrichment causes a lack of
    metal-poor stars (``G-dwarf problem''\footnote{The term ``G-dwarf
    problem'' describes the fact that the fraction of low-metallicity
    stars in the Milky Way disk is substantially lower than predicted
    by simple 1-zone models.}, see Section \ref{components}), similar
    to that observed in the Milky Way.
  
    We find a global age-metallicity relation and also a [O/Fe] to
    [Fe/H] relation, which are consistent with Milky Way observations.
    These relations are much more influenced by the choice of stellar
    yields, resp.\ the star formation history, than by the details of
    the galactic model, and have been found in previous galactic
    models as well \citep[e.g.][]{steinmetz_94, timmes_95,
    samland_98}. However, the pre-enrichment of the galactic disk,
    explaining the so-called ``G-dwarf problem'' and the sub-solar
    abundances in the outer halo gas are a result of dynamical
    processes and are therefore strongly influenced by the galactic
    model. The same is true for relations connecting the kinematics
    and chemical composition of stars; see Section \ref{intrels}.
 
\begin{figure}
      \resizebox{\hsize}{!}{\includegraphics{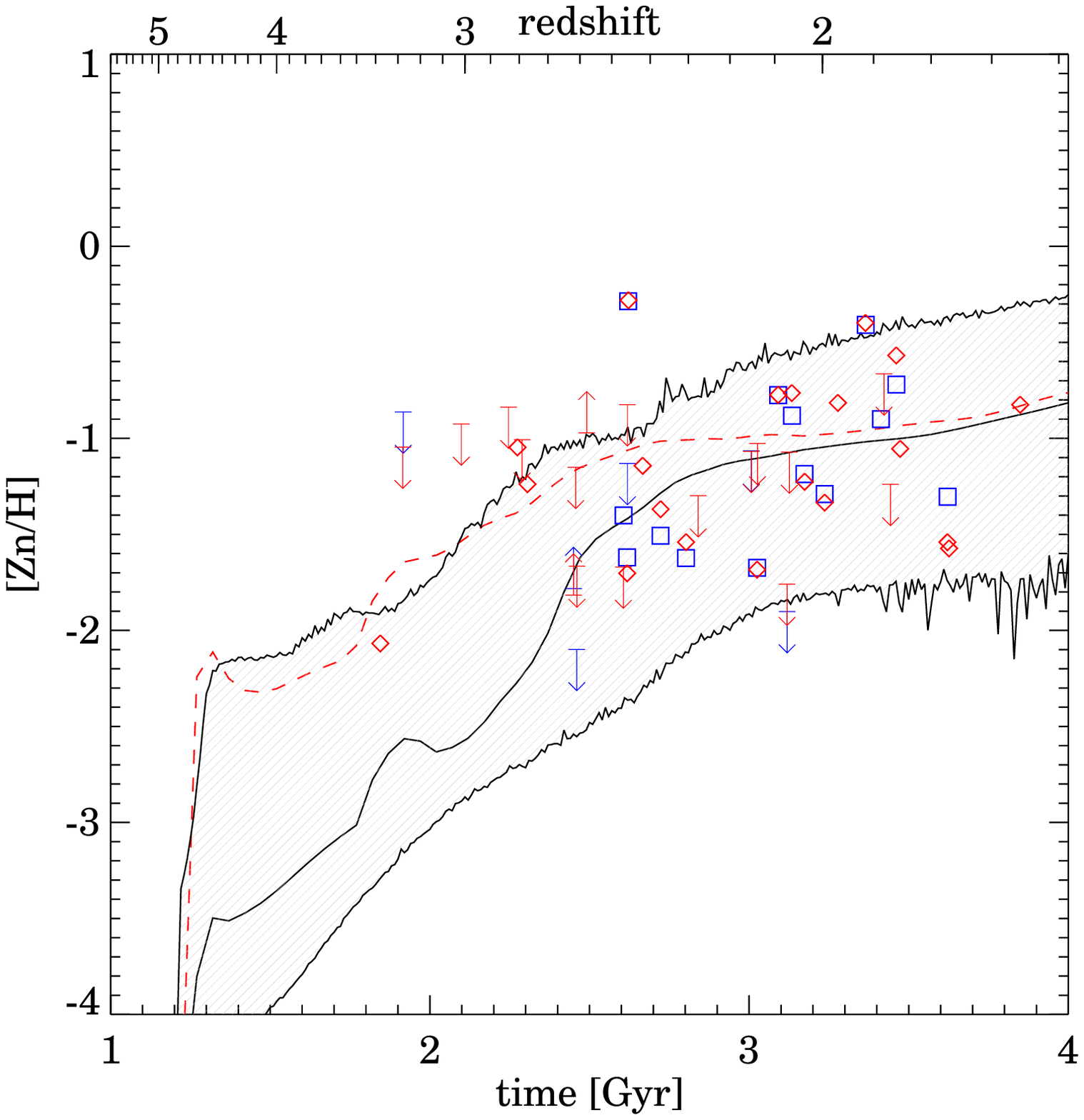}}
      \caption{Redshift evolution of the average zinc abundance [Zn/H]
               of the cloudy medium (full line) inside a
               galactocentric radius of $20 \kpc$. The shaded area
               shows the range of [Zn/H] metallicities in the model
               over this region. The points are data for damped
               \lymanalpha systems from \citet{pettini_99} and
               \citet{prochaska_01}, plotted as squares and rhombi,
               respectively.  The general trend of the model curve is
               consistent with the observed damped \lymanalpha
               abundances. The dashed line indicates the metallicity
               evolution of the hot gas component, which fills most of
               the galactic halo.}
      \label{MS2866f12}
    \end{figure}

    In Fig.~\ref{MS2866f12} we compare our results with the [Zn/H] data
    for damped \lymanalpha systems \citep{pettini_99, prochaska_01},
    using [Zn/H] because [Fe/H] may be depleted by dust. Damped
    \lymanalpha systems are an important test for galactic models.  If
    they are associated with forming disk galaxies, they can be used
    to measure the metallicity of the ISM of young, gas rich galaxies
    at high redshifts. The \lymanalpha systems on average show only a
    mild evolution in [Zn/H] for $z \approx 2-3$. Even though we
    consider only a single model with its specific evolutionary
    history, the predicted metallicity range as a function of redshift
    is consistent with the data. From the model we thus infer that the
    time needed to increase the disk metallicity to the typical values
    observed in \lymanalpha systems is $\sim 1 \gyr$. The shaded area
    in Fig.~\ref{MS2866f12} shows the range of metallicities from
    different locations in the model at redshifts $z > 1.6$.
    Fig.~\ref{MS2866f12} also shows the early [Zn/H] evolution of the
    hot ionized gas (dashed line). This should be representative for
    the average metallicity of metal-enriched clouds forming in the
    low density halo ISM, and thus for absorption of background
    quasars along halo lines-of-sight.

  \subsection{The ISM at $z = 0$}
 
    At $z = 0$, $5.6 \times 10^{10} \msun$ (15.7\%) of the initial
    baryonic mass of $3 \times 10^{11} \msun$ is still in the form of
    gas. The distribution of this late ISM is spatially more extended
    than that of the stars and the mass ratio of ISM to stars
    increases with distance from the galactic centre. Inside $10
    \kpc$, only about 2\% of the baryonic mass is still in the ISM,
    but inside $60 \kpc$ the fraction is already 43\%. 75\% of the ISM
    mass is in the cold and warm phases (molecular and atomic gas),
    but a significant fraction of 25\% is hot and ionized.  This hot
    gas extends far out into the halo and fills most of the halo
    volume (see Fig.~\ref{MS2866f8}). As discussed in the previous
    subsection, the hot gas is the chemically least enriched component
    at late times.

  \section{Formation of galactic components}
  \label{components}
 
    The model presented in this paper describes the formation of a
    massive disk galaxy with a total luminous mass of $2 \times
    10^{11} \msun$ inside $20 \kpc$ and a disk rotation velocity of
    $270 \kms$ at a radius $10 \kpc$, about 2.5 times more massive
    than the Milky Way. The galaxy model forms from inside-out
    radially and from halo to disk vertically. In Fig.~\ref{MS2866f6}
    one can clearly see halo, bulge, and disk components, which as
    described in Section \ref{history} form sequentially in this
    order. The model contains the full spatially resolved dynamical
    and chemical information for all of these components. With this
    information it is possible to study the detailed consequences for
    the stellar kinematics and abundances, that are implied by the
    assumptions made for the galaxy formation scenario and the
    modelling of the various physical processes involved, and to
    compare to a variety of observational data.

\begin{figure*}
     \resizebox{\hsize}{!}{\includegraphics{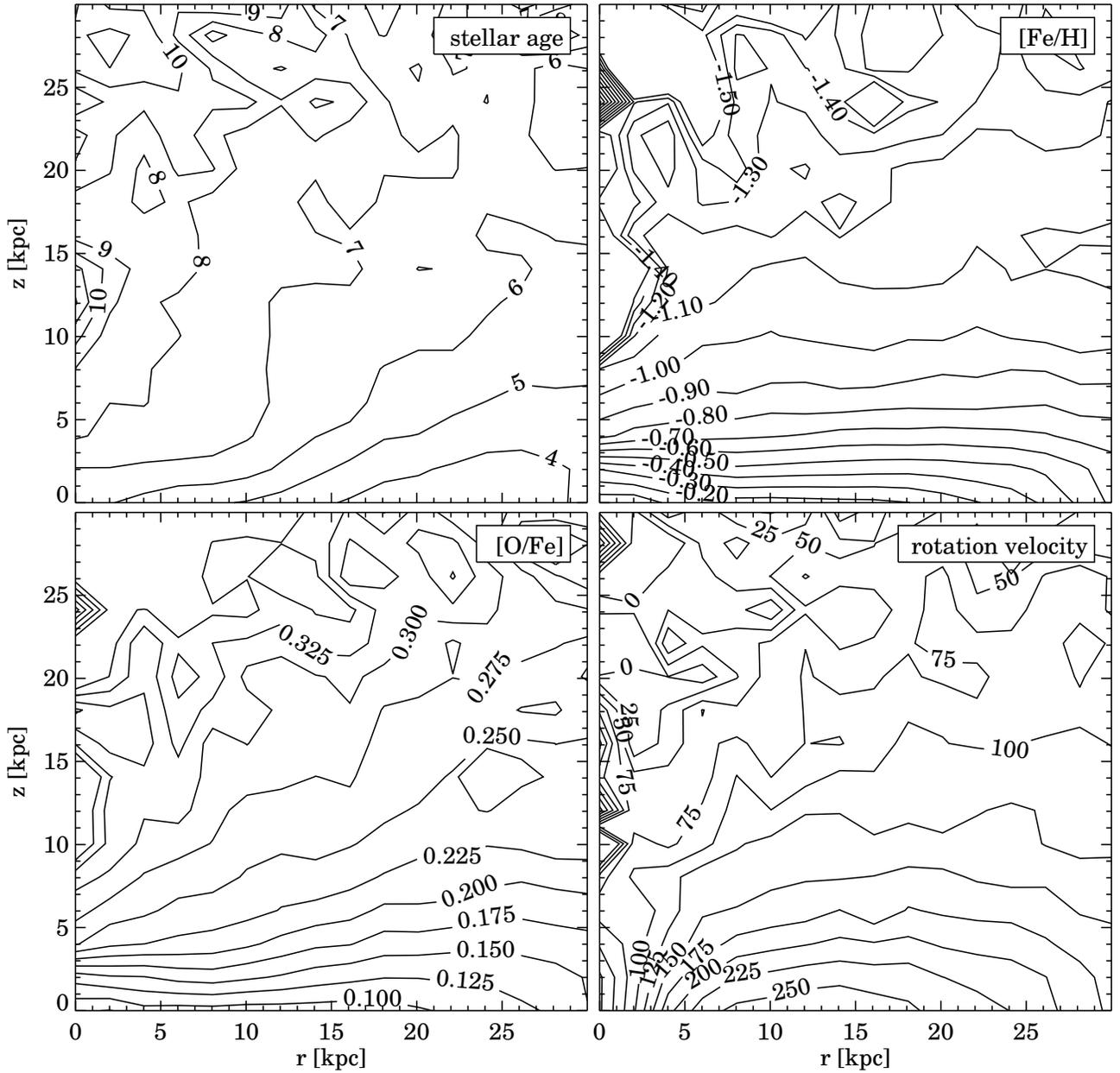}}
     \caption{The panels show the mean stellar age, [Fe/H], [O/Fe],
              and mean rotation velocity $V$ of the model stars at
              redshift $z = 0$, in the meridional plane as a function
              of radial ($R$) and vertical distance ($Z$) from the
              galaxy's centre. The mean stellar ages are given in Gyr,
              and the units of the velocities are $\pcmyr = 0.98
              \kms$.}
      \label{MS2866f13}
    \end{figure*}

    Figure \ref{MS2866f13} shows {\sl average} properties of the model
   stars at redshift $z = 0$: age, metallicity, [O/Fe], and rotational
   velocity.  These vary continuously as a function of position in the
   meridional plane ($R,Z$), in a way that reflects the interplay
   between star formation and enrichment on the one hand, and the
   vertically top-down and radially inside-out dynamical formation on
   the other hand.  Thus, the oldest stars can be found at low radial
   distances in the halo and the youngest stellar populations are
   located in the outer galactic disk. Because mean age and mean
   metallicity vary as a function of position, global mean values for
   age or metallicity are not sufficient to characterize the stellar
   population of a galaxy. Notice also that stars at a given mean age
   or metallicity can be found in a variety of positions.
 
\begin{figure}
      \resizebox{\hsize}{!}{\includegraphics{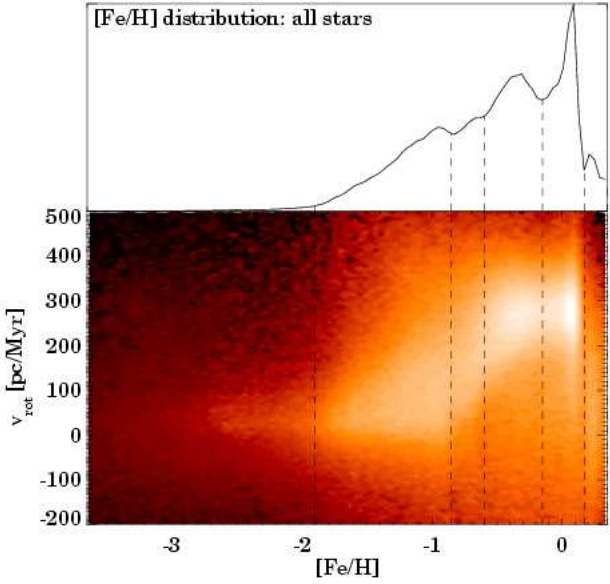}}
      \caption{Lower panel: Distribution of model stars in the
               metallicity--rotation velocity plane. Rotation
               velocities are perpendicular to the total angular
               momentum vector and are in $\pcmyr$. Upper panel: total
               metallicity distribution of all model stars, projecting
               along the rotation velocity axis. The dashed lines
               separate the subpopulations discussed in the text.}
      \label{MS2866f14}
    \end{figure}

    In Fig.~\ref{MS2866f14} we plot the complete distribution of model
    stars at redshift $z = 0$ in the plane of rotational velocity
    $\vrot$ versus metallicity [Fe/H]. This shows the expected
    evolution from non-rotating, low-metallicity stars formed early in
    the collapse to a rotating disk-like, metal-rich population formed
    at late times. However, this evolution is by no means described by
    a single dependence of $\vrot$ on [Fe/H]; rather there is large
    scatter in $\vrot$ at given metallicity, especially at low [Fe/H]
    values. In addition, there are also at least two concentrations of
    stars visible, at [Fe/H] $\simeq -0.4$ and [Fe/H] $\simeq 0.1$.
 
    The top panel of Fig.~\ref{MS2866f14} shows a [Fe/H] histogram of
    all stars, projecting along the $\vrot$ axis. Based on this
    histogram, we will now study the stellar populations in the
    different parts of Fig.~\ref{MS2866f14} separated by the dashed
    lines. In first order these might be described as extreme halo,
    inner halo, metal-weak thick disk, thick disk, thin disk, and
    central bulge components, somewhat reminiscent of the similar
    components known in the Milky Way. However, we do not intend to
    imply a detailed correspondence with the Milky Way components;
    there are significant differences with the present model. Recall
    that the dark and baryonic masses of this model are larger than in
    the Milky Way, and especially the dissipation parameter of the
    model may not be typical for the Milky Way.  Nonetheless, studying
    these components is useful to understand the course of the
    dynamical evolution and chemical enrichment in dissipative,
    star-forming collapse, and we will for definiteness use the Milky
    Way terminology in the following discussion of the properties of
    these populations.

  \subsection{[Fe/H]$<-1.9$ (``Extreme halo'')}
  \label{extremehalo}
  
    This component comprises all stars with [Fe/H] $<-1.9$; these
    stars form before the collapse and simultaneous spin-up of the gas
    goes under way. As the left row of Figure~\ref{MS2866f15a} shows,
    these stars form mostly in the first two Gyr after the start of
    the simulation (at time $1.2 \gyr$), but there is a small ongoing
    star formation in the halo even to late times from infalling
    low-metallicity gas (remember that we have not imposed a density
    threshold for star formation in this model). These stars have
    [O/Fe] near 0.4 typical for enrichment by SNe II. Their spatial
    distribution is nearly spherical, they rotate slowly if at all,
    and they have an asymmetric, broad distribution of eccentricities
    biased towards radial orbits. Here we have defined eccentricity
    as $e=(r_{\rm max}-r_{\rm min})/(r_{\rm max}+r_{\rm min})$ where
    $r_{\rm max}$ and $r_{\rm min}$ are the maximum and minimal
    three-dimensional radii of the orbit. The relative lack of
    low-eccentricity orbits at these metallicities (see Section
    \ref{intrels}) may be due to our neglect of small-scale structure
    in the initial conditions; compare with \citet{bekki_01}.

  \subsection{$-1.9<$[Fe/H]$<-0.85$ (``Inner halo'')}
  \label{innerhalo}

    Stars in this component have a metallicity range $-1.9<$ [Fe/H]
    $<-0.85$, covering the spin-up phase until the first minimum in
    the [Fe/H] histogram in Fig.~\ref{MS2866f14}. These stars form a
    more flattened spheroidal distribution with a flat density
    profile. They rotate with $\sim 70 \kms$, and the eccentricity
    distribution is less radially biased than for the extreme halo
    component. This component forms partly during the phase of rapid
    increase in the global star formation rate (Fig.~\ref{MS2866f6}),
    but there is also a younger part forming until late times. Here
    the selection in metallicity is important: the older part of these
    stars forms at relatively small radii when the metallicity there
    reaches the selected range, while the younger component forms at
    progressively larger radii from infalling low-density gas mixed
    with outflowing enriched material. This younger component may not
    form in real galaxies if stars can only form above a density
    threshold, as inferred from observations by
    \citet{kennicutt_98}. Further simulations also showed that a
    larger dissipation parameter (see Section \ref{dissipation}) 
    leads to a faster settling of the clouds into the disk and this
    suppresses the star formation in the halo at late times.

  \subsection{$-0.85<$[Fe/H]$<-0.6$ (``Metal-weak thick disk'')}
  \label{mwthickdisk}

    These stars with $-0.85<$[Fe/H]$<-0.6$ are intermediate in their
    rotational properties between a halo and a disk population (see
    Fig.~\ref{MS2866f14}). As Fig.~\ref{MS2866f15a} shows their spatial
    distribution is already strongly flattened. They have a sizeable
    rotation with peak of the distribution at $\vrot \simeq 150 \kms$,
    around 50-60\% of the circular velocity. The eccentricity
    distribution is broad and centred around 0.5. The oldest stars of
    this component form at around the peak of the global star
    formation rate when the gas still has [O/Fe] $\simeq 0.3$. A
    slightly larger fraction of stars in this range of [Fe/H] forms
    continuously until late times, at progressively larger radii,
    again favoured by the lack of star formation threshold. These
    stars are already enriched by SN Ia, as visible in their [O/Fe]
    distribution.

  \subsection{$-0.6<$ [Fe/H] $<-0.15$ (``Thick disk'')}
  \label{thickdisk}

    This component with $-0.6<$[Fe/H]$<-0.15$ is one of the distinct
    components in Fig.~\ref{MS2866f14}, and is clearly dominated by a
    disk component in its rotational properties and eccentricity
    distribution. However, Fig.~\ref{MS2866f14} shows that it also
    contains intermediate metallicity retrograde stars, showing that
    this metallicity selection also includes part of the bulge. These
    bulge stars cause the peak in the age distribution at $\sim 5
    \gyr$, with high [O/Fe]. The disk part has a relatively narrow
    distribution in [O/Fe] centred around $\sim 0.13$.

  \subsection{$-0.15<$[Fe/H]$<0.17$ (``Thin disk'')}
  \label{thindisk}

    The thin disk component, selected in the interval $-0.15<$ [Fe/H]
    $<0.17$, is characterized by near-circular orbits and a high
    rotation velocity, about $270 \kms$. This component forms with an
    approximately constant star formation rate and has mostly
    near-solar [O/Fe].  As in the previous thick disk component, a
    contribution from bulge stars is included by the metallicity
    selection. The thin disk shows the bar and spiral arms clearest;
    see the top panels in Fig.~\ref{MS2866f15a}. In time, it forms from
    inside out due to the accretion of material with progressively
    larger angular momentum.

    The thin disk shows only a shallow radial metallicity gradient of
    $0.02$~dex/kpc, which is due to efficient mixing of the disk ISM
    by the galactic bar. Beyond $16 \kpc$, where the mixing is not
    very efficient, a steeper metallicity gradient of $0.07$~dex/kpc
    is established. See also \citet{martinet_97} who found strong
    metallicity gradients only in non-barred galaxies. The metal
    production by the bulge and the mixing by the bar lead to a
    pronounced pre-enrichment of the disk, which is visible in
    Fig.~\ref{MS2866f16} by the lack of metal-poor stars.

    \begin{figure*}
      \begin{center}
        \resizebox{0.79\hsize}{!}{\includegraphics{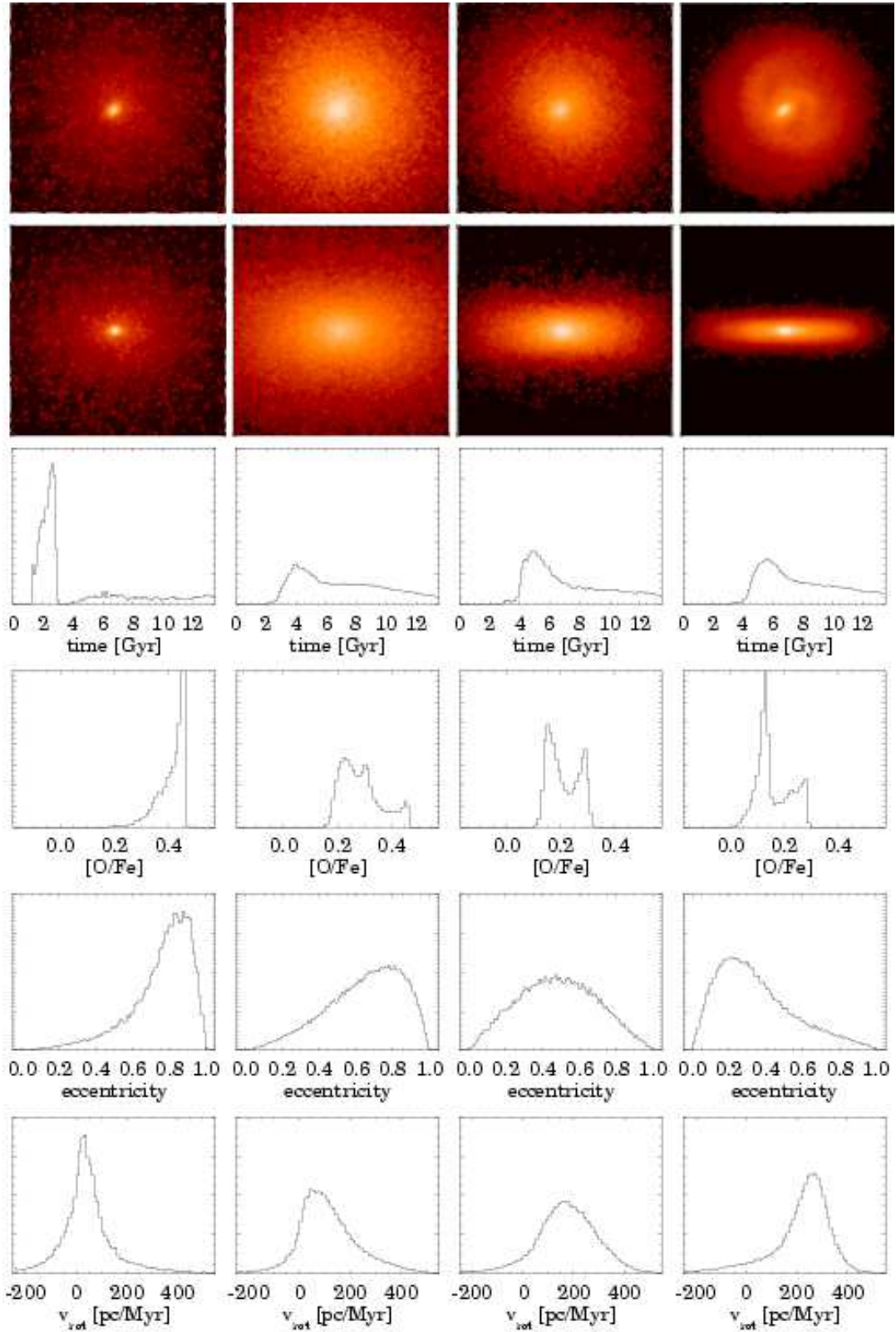}}
        \caption{Properties of model subcomponent selected by
                 metallicity according to the dashed lines in
                 Fig.~\ref{MS2866f14}. Columns from left to right show
                 all stars with [Fe/H] $<-1.9$ (``Extreme halo''),
                 $-1.9<$ [Fe/H] $<-0.85$ (``Inner halo''), $-0.85<$
                 [Fe/H] $<-0.6$ (``Metal-weak thick disk''), $-0.6<$
                 [Fe/H] $<-0.15$ (``Thick disk''), $-0.15<$ [Fe/H]
                 $<0.17$ (``Thin disk''), and [Fe/H] $>0.17$ (``Inner
                 bulge''). For each of these components, the panels
                 from top to bottom show the face-on projection onto
                 the disk plane, edge-on projection, distribution of
                 formation times, [O/Fe] distribution, eccentricity
                 distribution, and distribution of rotation
                 velocities. The top two panels in each row show an
                 area of $40 \kpc \times 40 \kpc$. Star formation in
                 the model starts at time $1.2 \gyr$.}
        \label{MS2866f15a}
      \end{center}
    \end{figure*}

    \addtocounter{figure}{-1}
    \begin{figure}
      \begin{center}
        \resizebox{0.79\hsize}{!}{\includegraphics{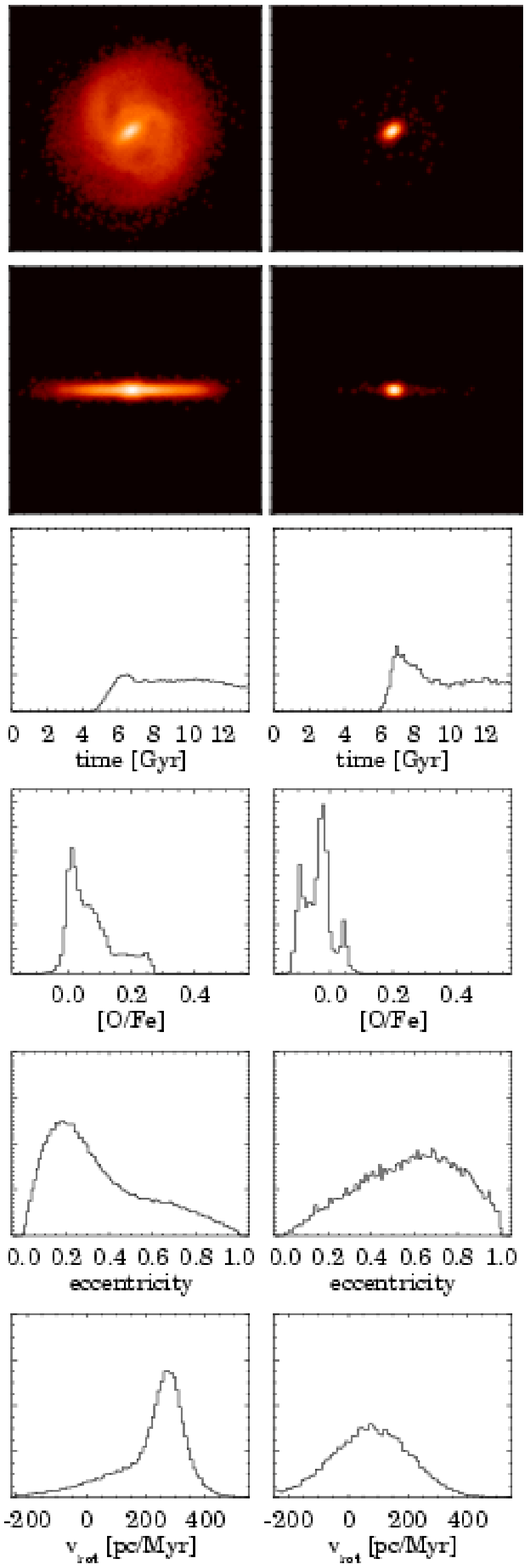}}
        \caption{ (continued)}
        \label{MS2866f15b}
      \end{center}
    \end{figure}

    \begin{figure}
      \resizebox{\hsize}{!}{\includegraphics{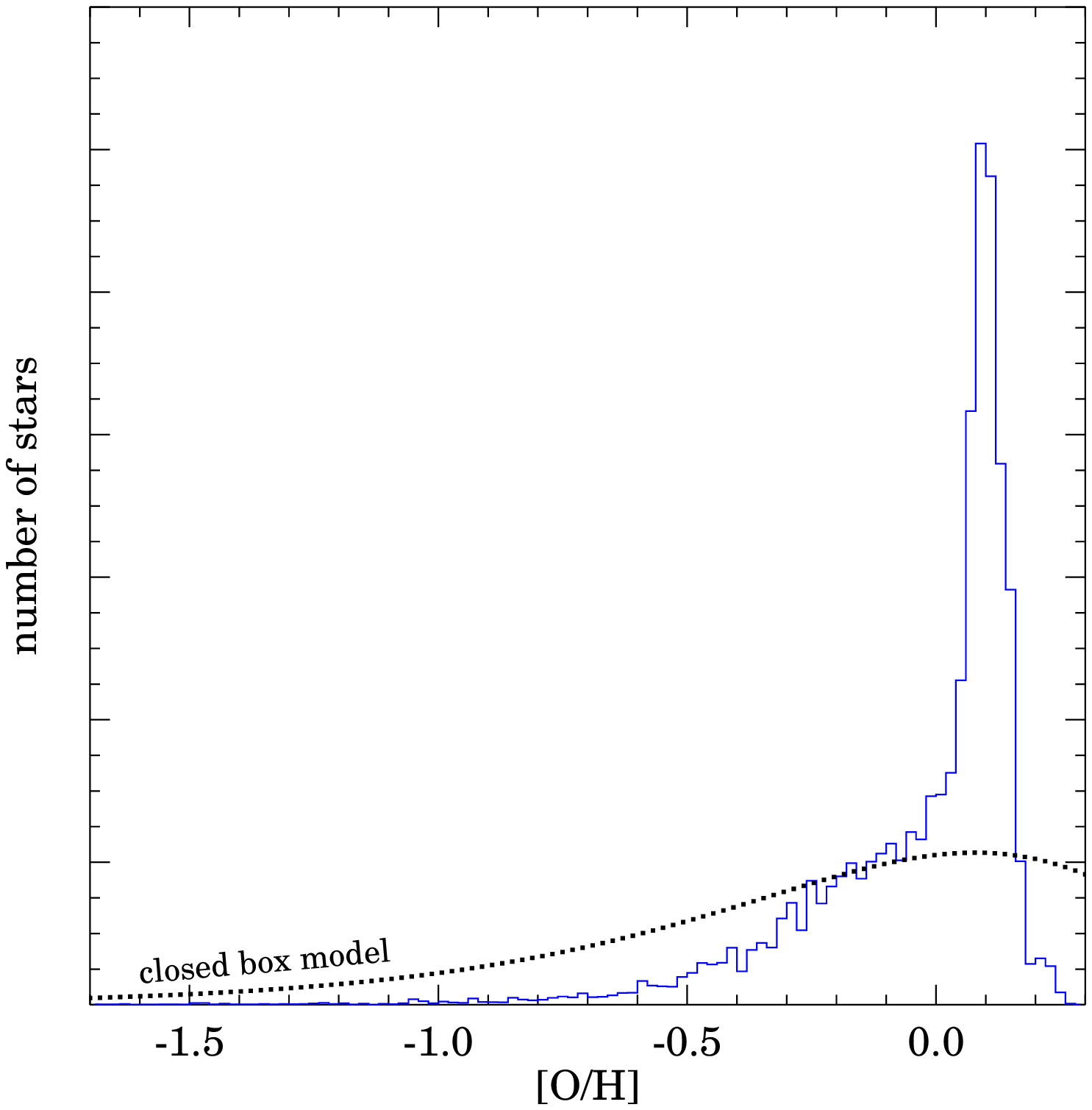}}
      \caption{Distribution function of oxygen abundances of G-dwarfs,
               selected by mass, in the model galaxy at redshift
               $z=0$. All G-dwarfs in the equatorial plane with
               galactocentric distances between $9.5 \kpc$ and $10.5
               \kpc$ are plotted. The dotted curve shows predictions
               of an instantaneous simple model with an effective
               yield $Y_{\rm eff} = Z_\odot$.}
      \label{MS2866f16}
    \end{figure}

  \subsection{[Fe/H]$>0.17$ (``Inner bulge'')}
  \label{innerbulge}

    The final metallicity component with [Fe/H] $>0.17$ consists
    mainly of stars in the central bulge. Notice that these most
    metal-rich stars do not form preferentially at late times; their
    age distribution is similar to that of the thick disk
    component. As the lower panels of Fig.~\ref{MS2866f15a} show, the
    metal rich inner bulge forms from $z \simeq 1$ in this model,
    rotates relatively slowly and has a broad eccentricity
    distribution.

    The total bulge population includes this inner component and a
    second, more metal-poor component included in the metallicity
    range of the ``thick disk''. Two subpopulations in the bulge is
    consistent with the observations of \citet{prugniel_01}, who find
    that galactic bulges in general consist of two different stellar
    components, an early collapse population and a population that
    formed later out of accreted disk mass.

  \subsection{Kinematics-abundance relations}
  \label{intrels}

    The previous subsections have shown that while it is useful to
    select stellar components by metallicity, such a selection is not
    ideal because it lacks spatial and kinematical
    information. Vice-versa, a purely kinematical selection would mix
    stars from halo and bulge, or metal-poor and metal-rich disk. A
    more physical separation would involve spatial, kinematic, and
    abundance information. This is possible with models such as this,
    because the complete population of model stars is available, but
    will be deferred to a later paper. Here, we will only show two
    relations between kinematic and abundance parameters for all stars
    in the model, which further illustrate some of its properties.

\begin{figure}
      \resizebox{\hsize}{!}{\includegraphics{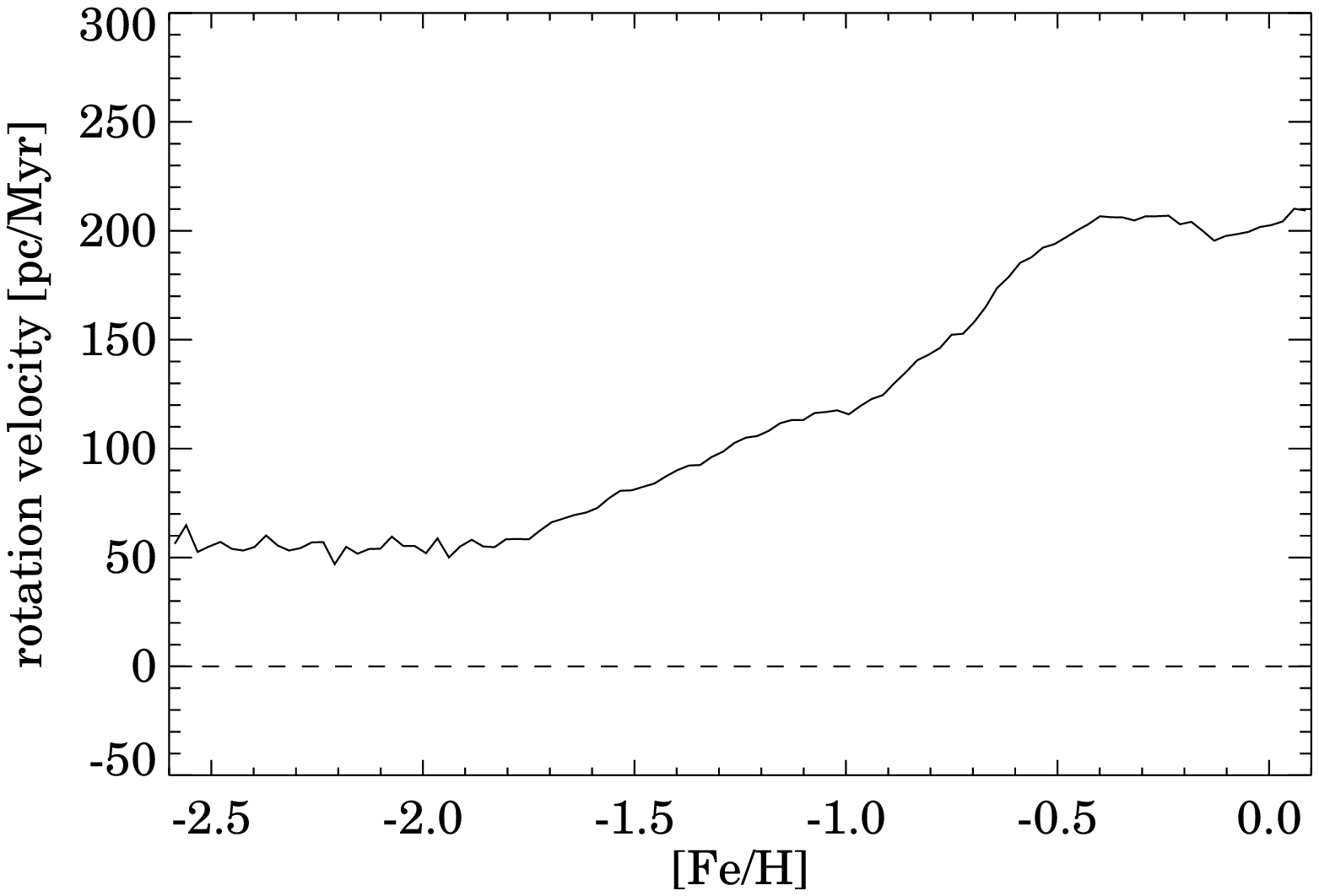}}
      \caption{Mean stellar rotation velocity as a function of [Fe/H]
               for all model stars.}
      \label{MS2866f17}
    \end{figure}

\begin{figure}
      \resizebox{\hsize}{!}{\includegraphics{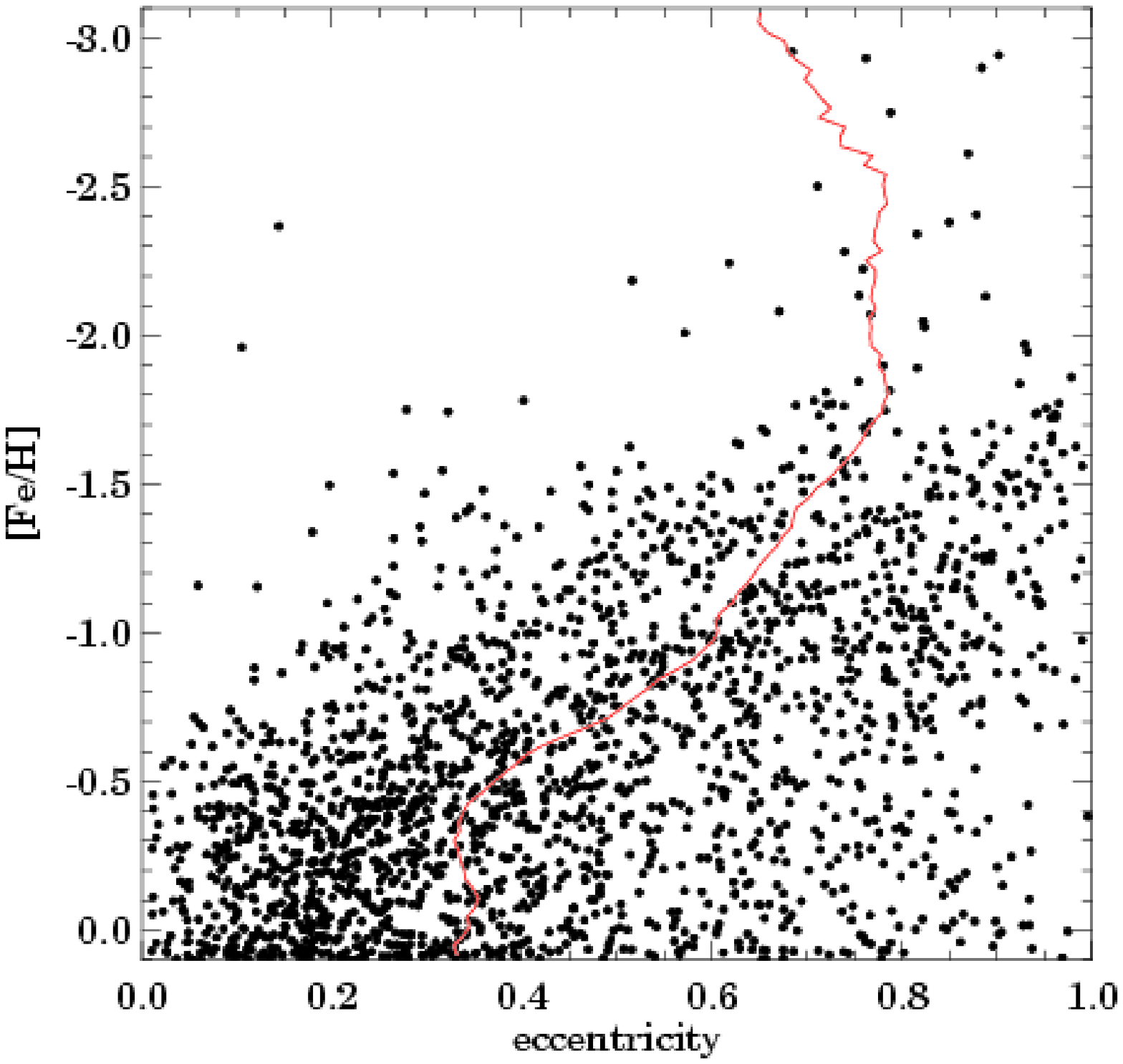}}
      \caption{Sample of stellar orbital eccentricities and mean
               eccentricity of all stars in the model, as a function
               of metallicity.}
      \label{MS2866f18}
    \end{figure}

    Figure \ref{MS2866f17} shows the mean stellar rotation velocity as a
    function of metallicity [Fe/H] for all model stars. For [Fe/H]
    $<-2$ $\vrot$ is independent of metallicity, at about $50
    \kms$. It then rises roughly linearly with [Fe/H] until the disk
    rotation velocity is reached at [Fe/H] $\simeq 0.4$. As is
    apparent from Fig.~\ref{MS2866f14}, there is a large dispersion
    around the mean value except in the disk components.  Figure
    \ref{MS2866f18} shows the mean orbital eccentricity and a sample of
    individual stellar eccentricities (defined in Section
    \ref{extremehalo}) as a function of metallicity. It is clear that
    at any metallicity, there is a large scatter in the orbital
    shapes. One of the causes of this is the very substantial
    deepening of the gravitational potential during the prolonged
    infall, which also affect orbits of stars formed early in the
    collapse. A second effect is that the turbulent velocities in the
    cloud medium are substantial during the collapse and during
    periods of large star formation rate, so that the stars made from
    these clouds inherit significant initial velocities. Thus all
    eccentricity distributions shown in Fig.~\ref{MS2866f15a} are fairly
    broad. When the various distributions shown in that figure are
    added with the appropriate mass weighting, a broad distribution
    results.

    Both the dependence of $\vrot$ on metallicity in
    Fig.~\ref{MS2866f17} and that of eccentricity on [Fe/H] in
    Fig.~\ref{MS2866f18} are similar to those observed in the Milky Way
    near the Sun \citep{chiba_00}, even though our model was not made
    to match the Milky Way and does not in detail describe the Milky
    Way. The main conclusion we draw from diagrams like
    Figs.~\ref{MS2866f17},~\ref{MS2866f18} is that realistic dissipative
    collapses are much more complicated than the simple model put
    forward by \citet{eggen_62} forty years ago. Therefore, inferring
    a merger history from the lack of rotation gradient with
    metallicity and a broad eccentricity distribution in the Milky Way
    halo is premature, irrespective of the fact that a fraction of the
    halo is likely to have been formed out of merging fragments
    \citep{ibata_01, helmi_99, chiba_00}. The important question,
    about the relative importance of the disruption of small scale
    stellar fragments, merging of gaseous fragments, and star
    formation in smooth dissipative accretion, all within their dark
    matter halos, will need more detailed modelling and more data to
    answer. All of these processes are expected in current CDM galaxy
    formation theories. The present simulations show that feedback is
    important for the final properties of the stars formed.

  \section{Conclusions}
  \label{conclusions}
 
    We have presented a model for the dissipative formation of a large
    disk galaxy, inside a growing dark matter halo according to the
    currently favoured \LCDM cosmology with $\Omega_\Lambda$=0.7,
    $\Omega_0$=0.3, $\rm h_0$=0.7. The dark halo mass accretion
    history is taken from the large-scale structure formation
    simulations of the GIF-VIRGO consortium \citep{kauffmann_99}. We
    use a total dark matter mass of $1.5 \times 10^{12} \msun$, a
    total baryonic mass of $3 \times 10^{11} \msun$, a spin parameter
    $\lambda=0.05$, and an angular momentum profile similar to the
    universal profile found by \citet{bullock_01}.
 
    We model the dynamics of stars and of two phases of the
    interstellar medium, including a phenomenological description of
    star formation, the most important interaction processes between
    stars and the ISM, and the feedback from massive stars and
    SNe. The dynamical evolution of the stars is treated with a
    particle-mesh code. For the dynamical description of the cold
    molecular and hot ionized gas components we use a
    three-dimensional hydrodynamical grid code.  Linked to the N-body
    and hydrodynamical code is a Poisson solver for the self-gravity
    and an interaction network, including a description of star
    formation, the heating and enrichment of type Ia and II SNe, mass
    return by intermediate mass stars, the heating and cooling by
    radiation, dissipation in the cloudy medium, formation and
    evaporation of cold clouds in the hot intercloud medium.  We
    discuss the range of uncertainty in the parameters that enter this
    description. Because of the self-regulation of these processes
    that is quickly established, even large changes in these
    parameters only lead to moderate changes, about a factor of 2, in
    the cloud velocity dispersion, hot gas temperature and ratio of
    hot to cold gas density.
 
    The simulation starts at a redshift of $z = 4.85$ with a dark halo
    of $2.1 \times 10^{10}\msun$.  Inside the growing dark halo a disk
    galaxy forms, whose star formation rate reaches a maximum of $50
    \msunyr$ at redshift $z \approx 1$.  Our main results can be
    summarized as follows:
 
    1. The formation and evolution of a galaxy in a dissipative
    collapse scenario is much more complex than predicted by simple
    models. The energy release by SNe and massive stars (feedback)
    prevents the protogalaxy from a rapid collapse and delays the peak
    in the star formation to a redshift of $z \approx 1$. The
    prolonged formation time causes fundamental changes in the
    galactic shape, the kinematics of the stars, and the distribution
    of the heavy elements.
 
    2. The galaxy forms radially from inside-out and vertically from
    halo-to-disk. The dynamics of the collapse strongly influences the
    star formation and chemical enrichment history, by modifying the
    gas density, cooling time, and thus star formation rate. The on
    average youngest and oldest stellar populations are found in the
    outer disk and in the halo near the galactic rotation axis,
    respectively. As a function of metallicity, we have described a
    sequence of populations, reminiscent of the extreme halo, inner
    halo, metal-poor thick disk, thick disk, thin disk and inner bulge
    in the Milky Way.  The first galactic component that forms is the
    halo, followed by the bulge and the disk-halo transition region,
    and as the last component the disk forms.

    3. An interesting feature of this model is the formation of a
    gaseous ring at a redshift of $z \simeq 1$ which collapses and
    forms a galactic bar. The bar induces numerous changes of the
    galaxy's properties.  It alters the density structure of the disk,
    induces spiral arms, enhances the mixing of ISM, and flattens
    metallicity gradients in the disk. It also channels gas into the
    centre, leading to prolonged star formation in the bulge region.

    4. The bulge thus contains at least two stellar populations: An
    old population that formed during the proto-galactic collapse and
    a younger bar population. The distinguishing feature between these
    populations is the [$\alpha$/Fe] ratio, with the bar population
    reaching [$\alpha$/Fe] $< 0$.
 
    5. The disk is the youngest galactic component, characterized by
    an approximately uniform star formation rate at late times. Its
    metallicity is approximately solar and its stellar metallicity
    distribution shows a pronounced lack of low-metallicity stars
    compared to simple closed-box models, due to pre-enrichment of the
    disk ISM.
 
    6. Early in the collapse an approximately spherical halo forms.
    As the collapse proceeds, the dissipation leads to spin-up of the
    gas, and newly formed stars acquire increasing rotation speeds, as
    expected. However, the distribution of orbital eccentricities as a
    function of metallicity for halo stars has large scatter. As a
    result, this distribution and the mean rotation rate as a function
    of metallicity are not very different from those observed in the
    solar neighbourhood. This shows that early homogeneous collapse
    models are oversimplified, and that conclusions based on these
    models about the degree of merging involved in the formation of
    the Milky Way halo can be misleading.

    7. The metal enrichment history in this model is broadly
    consistent with the evolution of [Zn/H] metallicity in damped
    \lymanalpha systems.
 
    8. The most metal-rich stars form approximately $1 \gyr$ after the
    peak of the star formation, when the SN Ia rate is at its maximum.
    These stars have the lowest [$\alpha$/Fe] ratios and, on average,
    an age of $\sim 5 \gyr$ in this model.  The infall of gas and the
    mass return from old stars decreases the average metallicity in
    the inner galaxy at later times.
 
    This chemo-dynamical model provides kinematics and metallicities of
    individual stars, but also can be used to obtain colours,
    metallicities and velocity dispersions of integrated stellar
    populations. It can therefore be used as a tool to understand
    observations of distant spiral galaxies, and connect them with
    stellar data in the Milky Way. In addition, it provides gas phase
    metallicities and temperatures, and star formation rates, as a
    function of time or redshift. \citet{westera_02} have already used
    this information to predict the colour evolution of large spiral
    galaxies, and have compared with bulge colours in the Hubble Deep
    Field. Further ``observing'' of such models and comparing with
    distant galaxy observations may lead to valuable insights on the
    galaxy formation process.

    A table of the positions, velocities, and metallicities for all
    stars in the model can be obtained by request from the authors.

  \begin{acknowledgements}
    Most of this work was supported by the Swiss Nationalfond. We
    thank M.Pettini for sending us the updated list of Zn abundances
    in \lymanalpha systems, and M.~Steinmetz for a careful referee
    report. The simulations were performed at the Swiss Centre for
    Scientific Computing (CSCS) and the Computer Centre of the
    University of Basel.
  \end{acknowledgements}

\end{document}